*Review*

# Agentic AI Systems in Electrical Engineering: Current State-of-the-Art and Challenges

**Soham Ghosh [1,2]* and Gaurav Mittal [3]**

1 Independent Researcher, Overland Park, KS 66211, USA
2 IEEE IAS/PES Kansas City Section, KS 66211, USA
3 Independent Researcher, Overland Park, KS 66211, USA
* Correspondence: sghosh27@ieee.org

**Abstract**

Agentic AI systems have recently emerged as a critical and transformative approach in artificial intelligence, offering capabilities that extend far beyond traditional AI agents and contemporary generative AI models. This rapid evolution necessitates a clear conceptual and taxonomical understanding to differentiate this new paradigm. Our paper addresses this gap by providing a comprehensive review that establishes a precise definition and taxonomy for "agentic AI," with the aim of distinguishing it from previous AI paradigms. The concepts are gradually introduced, starting with a highlight of its diverse applications across the broader field of engineering. The paper then presents four detailed, state-of-the-art use-case applications within electrical engineering, a domain where the impact of agentic AI systems is expected to be particularly significant. The high impact of agentic AI systems in the field of electrical power systems is primarily driven by global trends toward clean energy transition and higher levels of grid automations, all of which create an environment where agentic AI can be readily deployed and effectively leveraged. These case studies demonstrate current and innovative state-of-the-art, ranging from an advanced agentic framework for streamlining complex power system studies and benchmarking to a novel agentic AI system developed for survival analysis of dynamic pricing strategies in battery swapping stations. Finally, robust deployment of these autonomous agents brings a unique set of challenges that are discussed in this manuscript through detailed failure mode investigations. From these findings, we derive actionable recommendations for the design and implementation of safe, reliable, and accountable agentic AI systems, offering a critical resource for researchers and practitioners.

## 1. Introduction

The field of artificial intelligence (AI) is undergoing a significant paradigm shift, moving from systems that passively generate content to more of autonomous systems that can reason and react, play, and act, to achieve complex goals [1, 2]. The evolution as being witnessed in the US and globally is bound to have profound implications for high-stakes domains such as engineering. The progression of this technological trend can be broadly traced through three key phases: traditional agent-based AI systems, generative AIs, and the recent emergence of agentic AI frameworks.

Historically, the concept of an AI agent has been foundational, and followed a traditional definition of an entity that recognizes its surroundings and acts upon it [3]. Early agent-based systems [4, 5] focused on coordination and had limited problem-solving capabilities often within the bounds of predefined and narrow operational contexts. While effective for specific tasks like distributed control or optimization, these agents lacked generative reasoning and adaptability to novel problems. The second phase of this evolution was catalyzed by the advent of Large Language Models (LLMs), which underpin Generative AI (GenAI), with GenAI systems like ChatGPT and Gemini demonstrating unprecedented ability to comprehend context and produce multimodal human-like text, codes, and other media. In an engineering context, GenAI served as a powerful assistant, capable of drafting codes, summarizing technical documents, or suggesting initial design parameters. However, GenAI is by its very design 'reactive', i.e., it responds to a prompt to complete a discrete task but does not possess its own goals, nor can it independently verify or execute its outputs in a real-world feedback loop, by comprehending a complex goal, dissecting and solving it.

This is where the current and the third evolutionary phase comes into play; the third phase is that of Agentic AI. This represents an evolutionary leap by integrating the reasoning capabilities of LLMs with the classical agent's ability to act. Agentic systems are not merely content generators; they are goal-oriented systems that can autonomously decompose a high-level objective into a sequence of executable sub-tasks. By its design, Agentic AI can plan, memorize, reflect on its performance, and interact with external tools, such as compilers, simulators, or web APIs, to execute its plan and adapt to new information. A seminal study in this area is that of Park et al. [6], who introduced the concept of "generative agents" in a simulated environment. These LLM-powered agents exhibited believable emergent social behavior by operating on a "Perceive, Plan, Retrieve, Reflect, and Act" cycle, thereby demonstrating their long-term memory and self-directed activity. This work provided the foundational architecture for more complex agentic systems. Subsequent research has focused on formalizing these architectures, with comprehensive surveys by Wang et al. [7] highlighting the critical components of agentic AI with internal intelligence (reasoning, reflection, memory) and external tool invocation.

In complex engineering workflows, an agentic system can manage the entire process rather than just a single step. For example, in integrated circuit (IC) design, a GenAI model might generate a plausible but unverified block of Verilog code. In contrast, an agentic system can write the code, execute a simulation using an industry-standard tool, parse the resulting log file, identify a timing violation or bug, reflect on the error, modify its original code, and re-run the simulation, iterating until the design specifications are met. This closed-loop feedback mechanism, which mimics the core workflow of a human engineer, has already been demonstrated in recent studies such as ASIC-Agent (2024) and Agentic-HLS (2024) [8], which apply agentic reasoning to full ASIC generation and high-level synthesis, respectively. It is under this context that we present subsection 1.1 to discuss the potential of agentic AI as a transformative technology in engineering design and applications, and subsection 1.2 to illustrate the manuscript contribution and structure.

### 1.1. *Motivation and background*

The rapid emergence of agentic AI has prompted a number of valuable surveys [9, 10, 11, 12], which were critically assessed to understand their strengths and limitations. In [9], for instance, the authors define agentic AI as autonomous systems that pursue complex, long-horizon goals in dynamic environments with minimal human oversight, clearly distinguishing them from traditional, classical, and generative AI. The paper provides a beneficial but broad and somewhat scattered overview of agentic AI, covering core technical foundations such as reinforcement learning, goal-oriented and modular architecture, adaptive control, and mechanisms for planning, memory, and tool use, together with training and evaluation practices and supporting software frameworks. It surveys applications across multiple non-aligned sectors, including healthcare, finance, manufacturing, and education, highlighting adaptability, multi-goal management, and real-world deployment, while analyzing engineering challenges like goal alignment, environmental adaptation, and resource constraints alongside ethical and governance concerns involving accountability, bias, transparency, and safety. The authors outline open research directions and a roadmap for scalable, aligned, and well-governed agentic AI, positioning the survey as a foundational reference for both technical development and policy discourse.

The second survey [10] presents a beneficial but a loosely formed high-level overview of agentic AI that spans multiple technical and application areas rather than developing a tightly focused perspective. It defines agentic AI in relation to traditional AI and language-model-based agents, then reviews core components such as planning, memory, tool use, learning paradigms, and interaction mechanisms, along with representative agent architectures and frameworks. It also catalogs a wide range of use cases across non-aligned sectors including healthcare, finance, manufacturing, and education, treating these largely as illustrative vignettes of what agentic systems might do in practice. Overall, the survey is useful as a broad entry point into space, but its coverage is scattered across disparate domains, which limits its depth on any specific agentic capability or design principle.

The third survey by Ali and Dornaika [11] present a conceptually rich and methodologically rigorous survey that reframes agentic AI through a dual-paradigm lens, distinguishing a symbolic/classical lineage grounded in explicit planning and cognitive architectures from a neural/generative lineage centered on LLM orchestration and multi-agent pipelines. Their key contribution is to expose "conceptual retrofitting" in prior work and to propose a clear taxonomy that maps architectures, domains (e.g., healthcare, finance, robotics), and governance issues to the appropriate paradigm, while also identifying the strategic importance of neuro-symbolic hybrids and paradigm-specific evaluation and policy needs. However, the survey is theory-heavy and dense, with extensive historical and taxonomic exposition that can obscure direct design guidance for practitioners, and its paradigm split, while clarifying, sometimes overstates the incompatibility of symbolic and neural methods relative to current hybrid engineering practice.

The fourth survey by Bandi et al. [12] offer a broad, integrative review of "agentic AI" that aggregates definitions, tools, architectures, applications, input–output modalities, evaluation methods, and challenges across 143 primary studies spanning LLM-based and non-LLM systems. The paper is especially useful as a catalog, systematically tabulating frameworks (e.g., LangChain, AutoGPT, AutoGen, MetaGPT), architectural components (planning, memory, reflection, tool use), task domains, IO patterns, and qualitative/quantitative metrics while highlighting recurring technical, coordination, ethical, and security issues. Yet this breadth comes at the cost of conceptual sharpness the notion of agentic AI remains relatively inclusive and operationally loose, many sectors (healthcare, finance, retail, smart cities, education) are treated as parallel vignettes rather than analytically connected, and the survey largely reports and classifies prior work rather than advancing a strong, unifying theoretical perspective on what makes agentic systems distinct beyond being autonomous, multi-step, and goal-directed.

What emerges from a critical review of these four surveys is that these studies have successfully established a foundational understanding of agentic architectures, components (e.g., planning, memory, tool use), and general capabilities. However, this existing literature primarily focuses on the underlying computer science, natural language processing challenges, or broad, cross-industry applications [12, 13].

A significant gap persists in literature: to date, there has not been a comprehensive study on the specific state-of-the-art applications, challenges, and opportunities of agentic AI within the field of "engineering". This gap is particularly wide for "electrical engineering", a discipline that holds one of the highest potentials for an agentic AI revolution. This is because from an applications standpoint, agentic AI solutions can be readily leveraged in streamlining clean energy transition and decarbonization, navigating regulatory complexities and interconnection bottlenecks, and in assessing and mitigating complex power grid cybersecurity vulnerabilities. As such, a recent report published in Medium [14] on this subject adequately highlights how these emerging engineering areas may readily leverage and benefit from agentic AI solutions such as '*energy mix optimization agents*' to optimize energy sources based on demand forecasts, carbon intensity, and cost efficiency or '*carbon accounting agents*' to automatically track and report emission with an objective to exceed regulatory emission thresholds. From a broader adoption standpoint, these complex, multi-stage, and tool-dependent workflows can make an impact in other closely related areas such as power systems engineering management, event detection and diagnostic, congestion forecasting and renewable curtailment management, all of which are also ideal environments for autonomous, goal-driven agents. Preliminary smart grid applications [15] have already demonstrated some of agentic AI systems' disruptive potential, yet a systematic review with concrete example use cases that connect agentic theory to practical electrical engineering implementation is absent. This manuscript aims to fill that void.

### *1.2. Manuscript contributions and structure*

This manuscript provides the first comprehensive review of agentic AI systems specifically tailored for the electrical and computer engineering domains. Our primary contribution is to bridge the gap between the theoretical architecture of agentic AI and its state-of-the-art practical application in electrical engineering, while also systematically identifying the domain-specific challenges to its deployment.

The remainder of this paper is structured as follows:

- Section 2 discusses the taxonomical difference between traditional AI agents, Generative AI, and the new paradigm of Agentic AI, establishing a clear conceptual framework. In this section, we further explore illustrative use cases across the broader field of engineering to provide context for an agent-driven transformation of design and operational workflows.
- Section 3 forms the core of this manuscript and provides a detailed outline of four state-of-the-art cases studies in the field of electrical engineering, beginning with an MCP based agentic AI framework for streamlining of power system studies and benchmarking, followed by an MCP based agentic AI framework for enhancement of substation illumination studies, an agentic AI framework to generate engineering bill of quantity (BoQ) based on request for pricing (RFQ) documents, and finally outlining an agentic AI system for survival analysis of battery swapping station pricing strategies. For each of these case studies, we detail the success factors, limitations, and lessons learned.
- Section 4 conducts a critical review of the significant challenges and practical considerations for deploying trustworthy agentic AI in engineering. This includes security risks, such as the adversarial spread of false information and cascading misinformation from LLM rewrites, as well as practical barriers to building trustworthiness, such as model ambiguity, vulnerability to tool injection, and the necessity of robust human-in-the-loop (HITL) governance. In addition to discussing the problems, we present the state-of-the art mitigation solutions.

- Section 5 concludes the paper, summarizing our findings and outlining key directions for future research.

## 2. Foundational Theory of Agentic AI and Comparison with AI Agents

As illustrated in Figure 1(a), search interest in 'Agentic AI' as recorded by Google Trends was virtually nonexistent prior to 2025 and began to gain traction only after the first quarter of 2025. Figure 1(b-d) shows the top metro from US, China, and UK linked with the search interest in 'Agentic AI', as these are the countries that stands in the forefront of agentic AI research, based on bulletin published by Stanford's Human-Centered Artificial Intelligence (HCAI) working group [16]. In comparison, search popularity for the keywords 'AI agents' and 'generative AI' existed for some time (approximately since 2023 to present (2025)), and thus the question arises as to what is the difference between these three apparently similar and distinct paradigms? It is important that formal taxonomical understanding between these three separate areas of artificial intelligence be formed, and such a task has been undertaken in this section. Such formal taxonomical understanding is essential for several different reasons, including:

- *Guiding industry and innovation strategies:* Companies and developers need to align their research and development strategies to the right class of technology, as needed based on their particular use case. For example, generative AI companies might pivot towards agentic orchestration, only if they understand the fundamental differences and requirements between them.
- *Avoiding conceptual ambiguity:* In common media, these terms are often used interchangeably, even though they refer to fundamentally different capabilities:
    - Generative AI focuses on content generation based on learned patterns (e.g., text, image, code).
    - AI agents involve goal-driven entities that can act within predefined boundaries, often automating specific workflows.
    - Agentic AI introduces a higher level of autonomy, reasoning, planning, and coordination across tools or environments.
- *Enabling rigorous research and benchmarking:* Benchmarking of generative AI based content generation vastly differs from those for multi-step autonomous action. Hence, formal taxonomy provides a structured lens to define the different problem spaces.

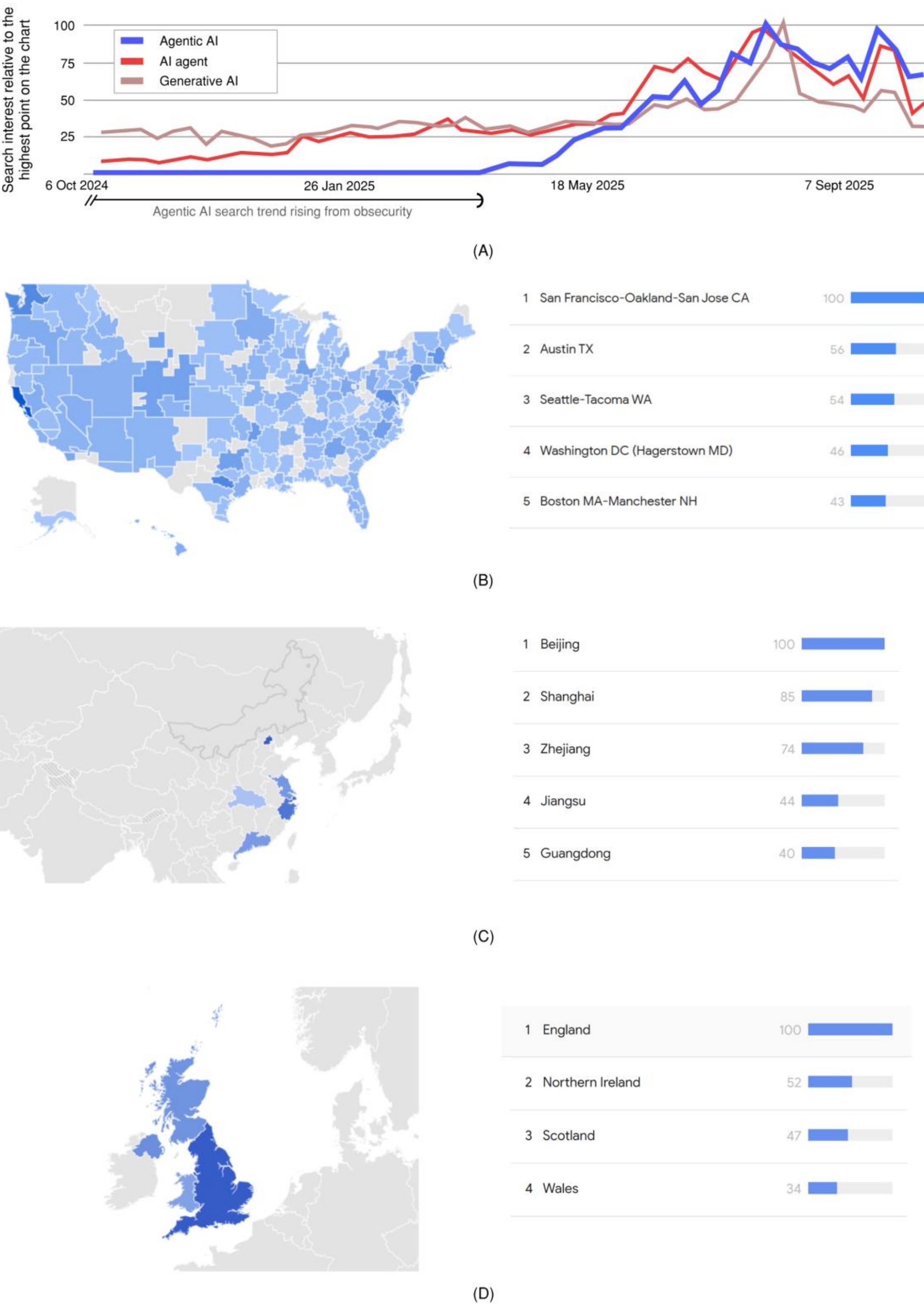

**Figure 1.** Trends in AI paradigms. **(a)** Google Trend showing 'Agentic AI' trend picking up post Q2 2025 from zero relative interest. **(b-d)** Top metro locations from US, China, and UK showing popularity with the search term 'Agentic AI' between Oct 2024 through Sep 2025.

Now that we have understood why formal taxonomical classification of these three areas of AI is essential, the focus is now shifted to form foundational understanding of these three areas. The taxonomical distinction is organized around how these systems start and control their work, how independently they can plan and act over time, and how many steps, domains, and tools they can coordinate to reach a goal. It also compares how they interact and learn (from simple prompt-and-response to ongoing adaptation), how they connect with people and other agents, and how their internal "wiring" (memory, architecture, and data access) supports more complex, long-running behaviors.

*2.1. Foundational understanding of AI agents*

AI agents are defined as an autonomous or semi-autonomous computational entity that reasons over information and takes prompt-based actions to achieve generally specific objectives [9]. It has a greater degree of adaptability and reasoning as compared to static, predefined sequences of instructions (e.g., shell scripts, Python scripts, macros) that execute exactly as written. They have found application within the domain of customer service [17], with bots that has the ability to showcase positive sentiment during customer service interactions [18] or providing a personal touch and improve the customer experience in customer service [19]. In industrial domains, multimodal retrieval-augmented generation agents are increasingly applied to accurately identify, count, and locate objects, particularly within complex scenes containing occlusions or small distracting objects [20], or AI meeting-summarizing agent automating the meeting note taking process in real time, ensuring that essential discussions contents are precisely captured and accessible for later review and usage [21].

*2.2. Foundational understanding of generative AI*

Generative AI refers to a class of artificial intelligence systems that are designed to create new content, such as text, images, audio, video, code, or other data, that resembles or extends human-produced work. Generative AI models are trained using large language models (LLMs) or large visual models (LVMs) that learn patterns from massive datasets of text, images, audio, or code. To produce an output, generative AI requires a user prompt (such as zero-shot, few-shot, chain-of-thought, tree-of-thought [22, 23]) and doesn't operate with autonomous goals. It must be highlighted that the multimodal capacity of generative AI allows systems to capture richer semantic relationships and expands the operational scope of AI systems beyond traditional unimodal applications. Such multimodal capacity is being widely seen in the field of generative AI [24, 25] (GPT-4 versus GPT-3, Switch-BERT versus BERT) and is especially powerful in engineering, design, and knowledge-intensive fields.

*2.3. Emergence and foundational understanding of Agentic AI*

So far, we have seen that AI agents perform well-defined tasks in structured environments, tasks such as process automation, customer assistance, etc., while generative AI gained the ability to understand and produce context-aware outputs across modalities (text, image, code, etc.). Agentic AI merges these two ideas, by bringing a higher degree of autonomy and planning (via goal decomposition) along with adaptive, multi-modal reasoning. An agentic AI framework usually consists of an orchestrator, a delegate, and a an ensemble of specialized AI agents, as shown in Figure 2, with the orchestrator agent analyzing the goal of the overall framework and decomposing the tasks into coarse- or fine-grained smaller decompositions [26], which are then passed onto the AI agents by the delegator. The delegator consolidates the results of the AI agents once their tasks (or subtasks) are completed and helps the agentic AI framework to move towards its goal.

Table 1 summarizes and further extends this section with a comprehensive set of distinctions between the three AI paradigms, namely agent-based AI, generative AI, and agentic AI.

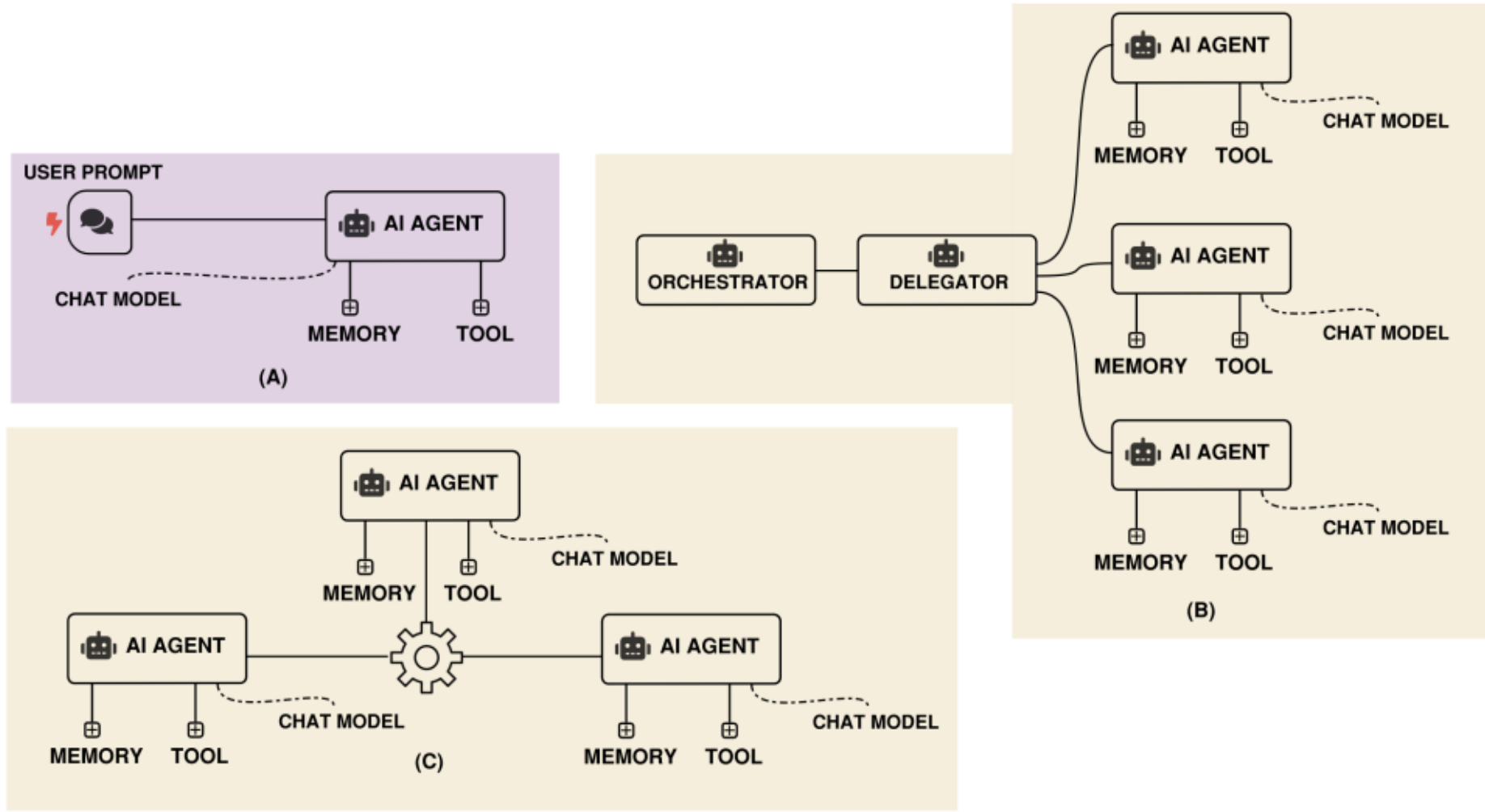

**Figure 2.** Overview of the framework architecture for: **(a)** AI agent. **(b)** Agentic AI system **(c)** Condensed representation of an agentic AI system.

**Table 1.** Comprehensive distinction between AI agent-based systems (agent-based AI, generative AI, and agentic AI).

| Distinction categories | Dimensions within the distinction categories | AI-agent | Generative AI | Agentic AI |
|---|---|---|---|---|
| **System initiation and control** | Initiation type | Rule-based or event-driven or prompt-driven | Prompt-driven | Self-initiating or goal-directed |
| | Autonomy level | Low | Medium | High — capable of autonomous planning and execution |
| | Task scope | Narrow, predefined | Broad but single pass | Multi-step, cross-domain workflow |
| **Interaction and coordination** | Learning | Minimal | Learned during pretraining; limited online learning | Online adaptation, reinforcement, and context-aware reasoning |
| | Coordination strategy | Static protocols and single agent | Minimal – single-turn interaction | Multi-agent orchestration, negotiation, adaptive coordination |
| | Social interaction | Limited | Human-centric | Multi-agent, human+agent+agent ecosystems |
| | Key roles | Prompt driven task execution | Content generation | Task decomposition, decision-making, dynamic collaboration, goal-based improvements |
| **Architectural distinction** | Control architecture | Single-loop or simple rule/LLM call | Single-stage model call | Hierarchical controller–executor pattern (orchestrator + specialist sub-agents), with explicit planning–acting–observing loops |
| | Execution topology | Primarily single-agent, point-to-point interaction with environment | Centralized model serving; client–server style interactions | Multi-agent topology (peer-to-peer or hub-and-spoke), supporting concurrent, partially independent agents with shared state |
| | Memory | Short term | Short term | Explicit use of shared memories, task queues, and world models to maintain long-running goals and cross-task state |
| **Data and core component** | External data access | Fixed APIs or statis data feeds | Limited to training data and prompt context | Dynamic access to APIs, tools, knowledge bases, and real-time systems |
| | Core component | LLM + tools | LLM + pattern selection (and no tools) | Multiple LLMs + tools |
| **Other examples** | - - - | Thermostat controllers | ChatGPT, Midjourney | Workflow orchestration; example, such as goal-oriented federated learning-based electric grid asset management to improve SAIDI, SAIFI metrics. |

### *2.4. AI Paradigms and their Broader Applications in the Field of Engineering and Technology Management*

Before introducing domain specific state-of-the-art use cases of agentic AI in the field of electrical engineering, a review of possible use cases of agentic AI in the broader domains of engineering and technology management should be discussed. From this broader standpoint, Table 2 may help the readers form a deeper understanding of use cases of AI based system. It illustrates the various use cases of these AI-based systems including AI agents, generative AI, and agentic AI systems.

It is important to note that while AI agents respond to user generated inquiries and generative AI synthesis new information, agentic AI systems are usually goal driven and encompasses a complex orchestration of tasks.

**Table 2.** Summary and comparison of state-of-the-art review papers on AI based systems (agent-based AI, generative AI, and agentic AI) for engineering applications.

| [Ref] Year | Engineering domain and agent type | Main points | Shortcomings |
|---|---|---|---|
| [27], [15] (2025) | Assistant for energy market rules, Agent-based AI (RAG) | RAG based AI agent **responding** to electricity market **inquiries** by referencing an internal database of market rules. | Concept could be expanded to contextual or structure-aware RAG for better retrieval performance of tabular or linked data. |
| [28] (2024) | LLM chemistry agent, Generative AI | Generative AI agent in the field of chemistry designed to successfully **plan and synthesize** new materials. | Strength of the in-built guardrails to check against controlled chemical and explosive depends on the exhaustive nature of the repositories being used. |
| [29] (2025) | Autonomous design of control algorithms, Generative AI | Using generative AI, successfully **evolved** a basic controller template into a high-performance controller. | Limited scope for evolution of control algorithms |
| [30] (2025) | Automotive engineering, Agentic AI | Robust Agentic AI framework with four AI agents for (a) styling, (b) CAD, (c) meshing, and (d) simulation having a **goal** to produce sophisticated aesthetic and aerodynamical designs. | Methodology for mesh evaluation remains limited. |
| [31] (2025) | 6G wireless network evolution Agentic AI | Agentic AI framework, designed with a **goal** to optimize the 6G services, communication, flow of data using context, semantics, and sustainability characteristics thus, realizing the true potential of 6G. | Agentic AI in a 6G framework requires high computation with the framework possibly suffering from latency issues in complex model scenario. |
| [32] (2025) | Effortless vehicular parking Agentic AI | Agentic AI framework aimed at improving urban mobility in densely populated areas. The agentic AI's **goal** is to use cooperative coordination between its agents to provide a frictionless parking experience for the user. | The evaluation is still laboratory-style. Area of improvement would be deployment in an actual urban mobility system or real parking infrastructure. |
| [33] (2025) | AI based urban planning Agentic AI | A fundamental reimagining of urbanism. An agentic AI framework that continuously recalibrates its operational priorities based on a set of **goals** for optimal urban outcomes. | Limited attention has been given to the necessary in-built guardrails that can secure equitable incentive distribution for marginalized urban communities. |
| [34] (2024) | Energy markets, Agentic AI | Agentic AI leveraging actor transformer-based critic (ATC) methodology having the **goal** for profit maximization through autonomous energy bidding and decision making. | A shared transformer encoder used for privacy-preserving purposes |

| [35] (2024) | Control system tuning/ evaluation Agentic AI | Agentic AI framework with a **goal** to perform complex control gain calculations and controller evaluations. | Current implementation is restricted to linear systems and conventional control strategies (PID and loop-shaping). |
|---|---|---|---|

Across the works summarized in Table 2, a clear progression can be observed from reactive, query-driven agents (RAG-based assistants), through content-generating systems (generative AI), to fully goal-directed agentic AI frameworks that orchestrate multiple tools and sub-tasks in complex engineering workflows. While the assistant-style agents in the energy market context primarily focus on accurate retrieval and explanation of existing rules, the generative AI chemistry and control-design applications demonstrate creative synthesis capabilities but remain largely single-agent and narrowly scoped around well-defined design spaces. In contrast, the agentic AI frameworks in automotive design, 6G network evolution, urban mobility, urban planning, and energy markets explicitly coordinate multiple specialized agents toward higher-level system goals (e.g., aerodynamic performance, quality-of-service in 6G, frictionless parking, equitable urban outcomes, or profit-maximizing market participation), yet they also expose new challenges around scalability, latency, evaluation, and governance that are not systematically compared in the existing literature.

Taken together, these studies highlight several open research directions that cut across domains. First, there is a methodological gap in how to rigorously form multi-agent, goal-driven systems in safety-critical settings such as automotive engineering, where security and verification guarantees are as important as raw performance. Second, there are unresolved tensions between autonomy and oversight: guardrail design in chemistry and urban planning, privacy-preserving mechanisms in market agents, and fairness in urban incentive structures are treated in a domain-specific manner, but a unified framework for safety, transparency, and accountability in agentic AI has yet to emerge. Third, most agentic AI case studies are still demonstrated in constrained or laboratory-style environments (e.g., simulated parking or simplified control architectures), leaving open the question of how these frameworks will behave when deployed at scale in pilot implementations or in real infrastructures with heterogeneous data, legacy systems, and human stakeholders. In the next two sections, sections 3 and 4, we will showcase some state-of-the-art deployment that goes beyond laboratory style environments along with failure mode investigation and recommendations to mitigate the failure modes.

## 3. Domain Specific State-of-the-Art Use Case Illustrations of Agentic AI Systems in Electrical Engineering Applications

At this point, we have developed clear taxonomical understanding of AI agents, generative AI, and agentic AI framework in section 2 and have seen the use cases of all three of these AI paradigms from a broader context of engineering and technology management. In this section, we are going to discuss in-depth several domain specific use case illustrations of agentic AI systems in the field of electrical engineering.

Before we can undertake discussions related to in-depth domain specific use cases, we need to form some understanding of the model context protocol (MCP), and how it differs from other established mechanisms. While established mechanisms such as function calling, ReAct-style reasoning–action loops, and multi-agent orchestration frameworks (e.g., AutoGen) provide effective strategies for tool invocation and agent coordination, MCP has gained rapid traction as a unifying interoperability layer. Unlike these approaches, which primarily define *how* models' reason, act, or collaborate within an application, MCP standardizes *how tools themselves are exposed, discovered, and accessed* across heterogeneous systems. Forming a clear distinction between these paradigms, as undertaken in Table 3, is therefore essential: conflating reasoning strategies, orchestration mechanics, and protocol-level interfaces obscures architectural responsibilities and limits reproducibility, portability, and ecosystem-level scalability of agentic AI systems.

**Table 3.** Conceptual comparison of function calling, ReAct reasoning, AutoGen orchestration mechanics, and the Model Context Protocol (MCP) across abstraction levels and system responsibilities.

| Dimension | MCP | Function calling | ReAct | AutoGen |
|---|---|---|---|---|
| "What problem does it solve?" | Tool *interoperability & portability* across many tools/systems | Reliable *structured tool invocation* from the model | Better *tool-using reasoning loop* (plan-act-observe) | *Multi-agent* collaboration + orchestration |
| Where it lives | Between agent app and external tool providers | In the model API contract (schemas → tool calls) | In the prompt + agent loop | In application framework/runtime |
| Tool discovery | Built-in (tools/list, updates) | Typically, static list you provide per request | N/A (uses whatever tools your loop provides) | Depends on the integration (can wrap tool calling / MCP) |
| Biggest win | Avoids separate N×M integrations; provides a "plug-in ecosystem" | Predictable JSON args; allows for schema adherence | Reduces hallucination via "act to verify" | Scales complex workflows via role-specialization |

Here it must be highlighted that up until 2024 the interaction between LLMs and external tools was fragmented and required bespoke plugins or proprietary APIs/ function calling. This limitation kept LLM based AI agent interaction brittle, siloed, and hard to scale or audit. The mode of communication between LLMs and external tools revolutionized in 2025, with Anthropic's introduction of the model context protocol (MCP) [36, 37], which was quickly adopted by the majority of the LLM powerhouses including OpenAI, Google, Meta, Microsoft, and Amazon. The wave of accessibility [38] offered through MCP's capability-oriented way for models to discover tools, and exchange structured inputs/outputs, accelerated innovation across industries, including the electrical engineering sector. [39] is an excellent resource for readers interested in forming a deeper understanding of MCP architecture, including its core components, different layers and communications, and the MCP server lifecycle, along with an outstanding collection of community driven MCP servers.

In the field of electrical engineering the impact was felt almost immediately, with developers creating custom MCP [40] thereby allowing an LLM based agentic system to orchestrate heterogeneous software components, such as power-flow solvers, contingency analysis and protection studies, electromagnetic transient (EMT) simulation modules, asset health services, along with office productivity applications. As an evolving alternative workflow, natural language-based goals can now be passed into an agentic AI framework, which decomposes the task and passes it to its AI agents which leverage MCPs to access powerful tools such as power system simulation or computer graphic software. Under this evolving framework, engineers can now specify a goal to an agentic AI system such as "*run benchmark an array of power system simulation software, and perform a power flow analysis and a harmonic study with and without filter banks using the best performing simulation software, draft end-results with simulation plots and a compliance checklist*", and allow the system to plan the work sequence, validates parameters against schemas, executes calls, tracks artifacts, and automatically recover from errors with structured retries. The added benefit is that with the intent now being declared in natural language rather than being hard-wired, instructions between specialist AI agents can be ported, provided valid MCPs exist. This not only reduces integration overhead but also makes agentic intelligence a practical, auditable solution for end-to-end engineering workflows.

For agentic AI workflows, the common starting point is to ramp two or three pilot projects driving them to organizational maturity level. Once such a level of maturity is reached, an organization might be able to compare a broader set of other use cases based on their business impacts and implementation complexity and decide on selecting the ones with the highest business impacts provided the tools and techniques to bridge the implementation complexity is within reach. Table 4 outlines a collection of such use cases, shortlisted in consultation to in-house AI strategists, and decomposed in terms of business score and implementation complexities. The use cases with the highest scores in both categories are deemed as suitable candidates for further development into an agentic AI framework under pilot implementation.

**Table 4.** Scoring breakdown was created in consultation with AI strategist to quantify the nine electrical engineering agentic AI use cases in terms of business impact and implementation complexity.

| Agentic AI use cases | Business impact | Implementation complexity | Overall score |
|---|---|---|---|
| Distribution level asset: event detection and diagnostic | Limited incremental business value; localized optimization potential in asset monitoring<br>**Score: 1/5** | Technically straightforward; mature algorithms exist; minimal integration overhead<br>**Score: 1/5** | **2/10** |
| Congestion forecasting and renewable curtailment management | Useful for operational visibility but limited direct customer monetization<br>**Score: 2/5** | With robust SCADA in modern power system data availability is fairly easy, low real-time control integration complexity<br>**Score: 1/5** | **3/10** |
| Renewable energy forecasting and planning | Indirect customer value via improved grid planning and operational foresight<br>**Score: 2/5** | Involves multi-timescale forecasting, meteorological inputs, and planning models<br>**Score: 3/5** | **5/10** |
| Substation SCADA alarm and management | Moderate operational benefit through reduced alarm fatigue and response time<br>**Score: 1.5/5** | Requires structured data ingestion from SCADA, basic agent coordination<br>**Score: 4/5** | **5.5/10** |
| PSPS for wildfire mitigation | Very high business value due to safety, regulatory, and reputational considerations<br>**Score: 4/5** | Operational triggers well-defined; relatively low integration and orchestration complexity<br>**Score: 2/5** | **6/10** |
| RFQ to BoQ for engineering services and EPC | High commercial value through automation of bid development and engineering cost reduction<br>**Score: 3.25/5** | Moderate to high complexity due to natural language to structured BoQ translation workflows<br>**Score: 3.25/5** | **6.5/10** |
| Complex substation studies | High value due to direct engineering productivity gains and reduced study turnaround time<br>**Score: 4/5** | Involves multiple software orchestration, agent coordination, and high-fidelity modeling<br>**Score: 4/5** | **8/10** |
| Power system studies and benchmarking | High system-level value - enables complex power system simulation-based studies to be executed<br>**Score: 4.5/5** | Requires a complex orchestration of multiple modeling environments<br>**Score: 3.75/5** | **8.25/10** |
| Survival analysis of EV pricing models | High value for strategic planning and business model forecasting. Adoption of the correct pricing model may mean survivability of an EV charging/swapping provider.<br>**Score: 5/5** | Advanced statistical knowledge required<br>**Score: 5/5** | **10/10** |

Based on the scoring from Table 4, the nine use cases are plotted in Figure 3 for ease of visualization, and the four highest scoring cases (deep blue) are developed further, with details of each of the agentic AI framework (along with limitations, lesson learned, and extensions) provided in sections 3.1 through 3.4.

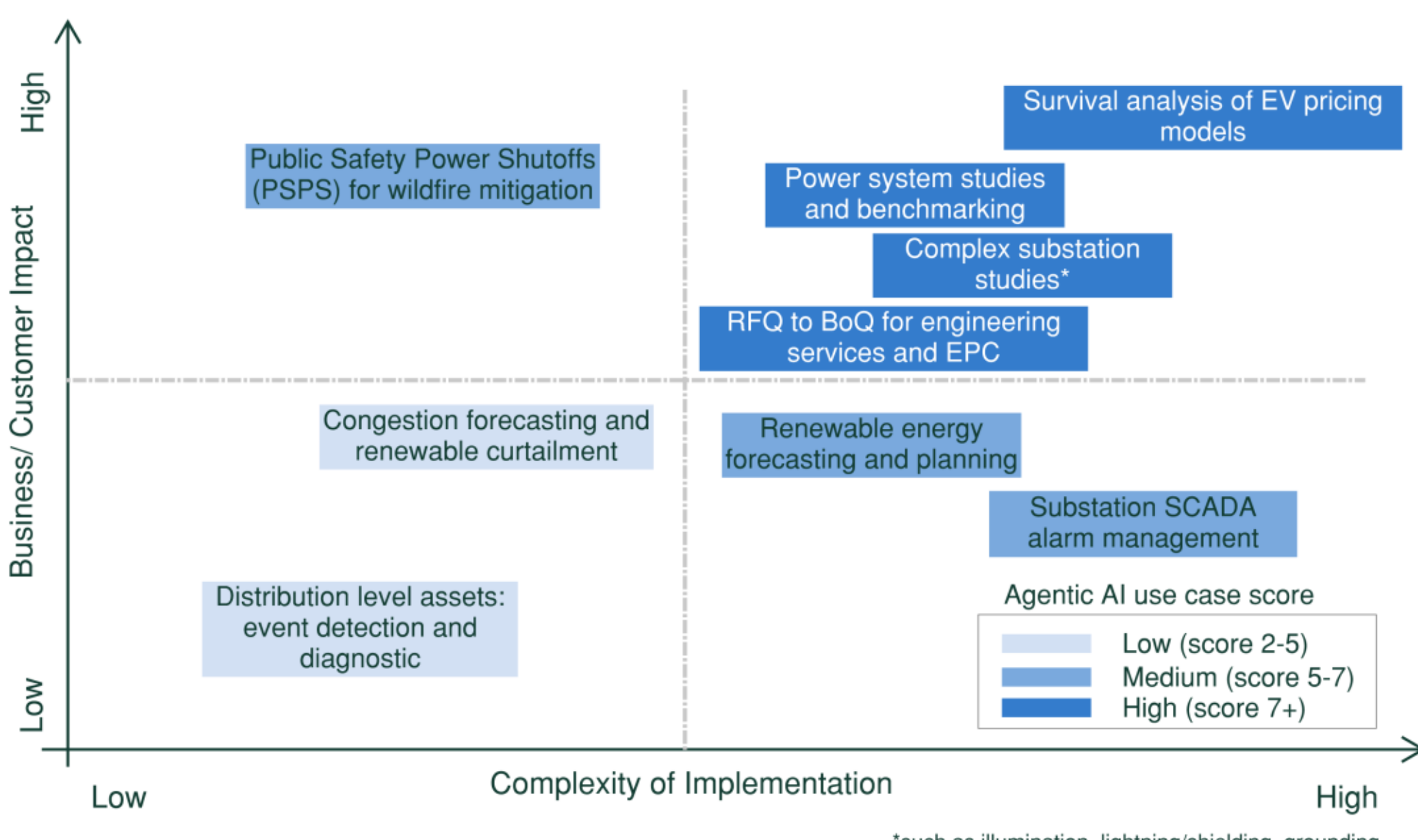

**Figure 3.** Prioritization guide for developing agentic AI frameworks while maximizing business impacts for hard to implement tasks.

### *3.1. Case study of MCP based agentic AI framework for streamlining of power system studies and benchmarking*

The first case study presents an agentic AI framework showcasing how a group of specialized AI agents having access to power system simulation tools via MCP be used for benchmarking of power system simulation software with the goal of assessing the best run times and computation accuracy. Traditionally, for power system studies such as power flow, contingency analysis, and transient simulation analysis, system engineers need to learn the intricacies of simulation tools such as *PSS®E, PowerWorld, DIgSILENT*. Often times one may have to leverage multiple software to accomplish a complex study objective, and mastering a full stack workflow takes years and often poses a huge entry barrier for early career professionals. With an agentic AI workflow, the engineer will not need to master the inner working of each of these programs and can focus on high level decision making and directing the agentic AI system to handle the program-interface level tasks. Based on its benchmarking assessment, the agentic AI framework may use the most relevant set of simulation software, based on simulation capability, run time, and accuracy. With an agentic flow such as the one proposed in Figure 4(a) such an objective can be achieved, with the agentic AI framework benchmarking the different software that is available to it, and subsequently decomposing a complex simulation task and leveraging the constituent AI agents as needed to provide the final outcome. A contrasting traditional and standalone *pandapower* workflow is shown in Figure 4(b) where a system user has to stepwise create a network model, compile code to run the desired power system studies, and manually document and assess the results. Depending on the objectives, one may have to orchestrate a full simulation study manually across multiple software to be able to compile final results. Figure 4(c) shows an improvement over Figure 4(b) with illustration of the prompts and the outputs of a standalone "*pandapower* AI agent". Instead of working inside the coding environment of the *pandapower* software, an user can leverage this AI agent and upload the desired network file and prompt the AI agent to execute a power flow, an N-1 contingency analysis, or a short circuit study, and generate the output in a desired reporting format.

As one may observe, an agentic AI framework poses much higher degree of autonomy and adaptability, compared to constituent AI agents. Given its goal-oriented nature, it can benchmark software performance and use these benchmarking results to intelligently invoke the most appropriate set of software (via orchestrator and delegator) to execute a complex power system study. In contrast, a standalone AI agent can perform a very limited set of tasks without any goal decomposition or multi-agent coordination capability. Critical evaluations of this case study are outlined as follows:

- **Success –** The power system agentic AI framework was able to successfully benchmark several power system studies, such as power flow, contingency analysis, etc. across *pandapower, PSS®E, PowerWorld.* Based on its benchmarking results and fulfilling its subsequent goal, it was able to intelligently decompose a multi-objective system study, selecting the most well-suited specialized agents, and ultimately generating a full study report.
- **Limitations –** At the time of writing this manuscript, none of the power system software providers offered standardized official MCPs, compliant with stable version of their software. Having standardized and stable versions of vendor provided MCP would allow for wider collaboration on these agentic workflows.
- **Lessons learned –** Depending on the nature of the prompt that was used, it was observed that at certain times, the agentic AI framework tried to read or rewrite the entire network file before passing it to the power system software for processing. When this happened, the LLM token limits were quickly reached, and subsequent steps were aborted. Refining the prompts used by individual AI agents to be more explicit usually helped in mitigating this behavior. For instance, instead of "read the file" as a prompt template, a sub-AI agent using a prompt template like "use the filesystem tool to read the contents of /path/to/your/file.txt" yielded much better results, and all operations were witnessed to be completed within the LLM token limits.
- **Extensions** – The framework can be extended for more complex benchmarking and power system studies. For example, the agentic AI framework can be further developed for DER interconnection screening, or other specialized studies such as distribution system loss/efficiency studies. Results from a constituent AI agent, such as the *OpenDSS* specialist agent maybe subsequently fed to a *PowerWorld* AI agent to perform an optimal power flow with qualifying DERs and generating the MW marginal costs.

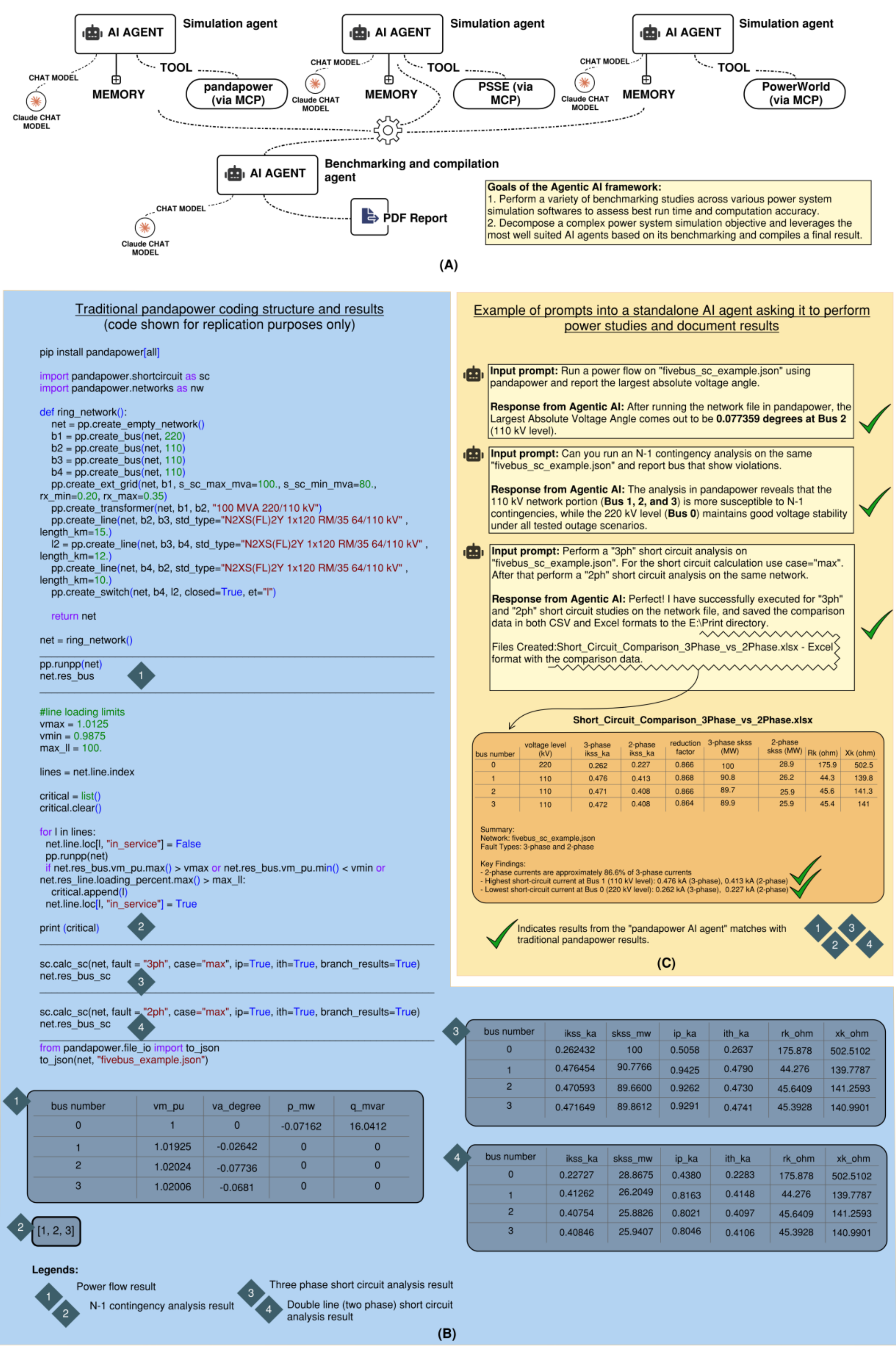

**Figure 4.** Agentic AI framework for power system benchmarking. **(a)** Detailed agentic AI framework with an ensemble of specialized AI agent and well-defined goal(s). **(b)** Native pandapower code structure shown for creating an electrical network and executing a power flow, a contingency

analysis, and short circuit studies. **(c)** An AI agent-based workflow with an user prompting into a standalone "pandapower AI agent" to invoke pandapower and have it perform power flow, contingency analysis, and short circuit analysis.

*3.2. Case study of MCP based agentic AI framework for enhancement of substation illumination studies*

The second case study presents an agentic AI framework, showcasing how a group of specialized AI agents can collaboratively automate and streamline an entire electrical substation illumination calculation and reporting workflow. An illumination study is a vital component of overall substation design, as there are industry guidelines, such as the National Electric Safety Code (NESC) section 111-1 [41], which necessitate certain levels of illumination levels that must be maintained within the substation yard and around the major electrical equipment to ensure safe operation and maintenance at night. Traditionally yard luminaires are placed on lightning masts, bus support steel structures or take off structures, within an illumination computation software such as *AGI32* or *Acuity Brand's Visual Lightning* by the substation design engineer. The quantity of the luminaires to be placed is usually determined by the overall footprint of the substation and the location of the major electrical equipment within. Usually, zoning restrictions apply, and care is taken to reduce any excessive bleeding of light outside the substation perimeter fence.

Our proposed agentic AI system facilitates this entire process, with the framework outlined in Figure 5(a). The process usually starts with defining a set of goals that the agentic AI framework should aim to achieve. With the general outline of a substation fed into the agentic AI framework, it can intelligently evaluate the major equipment, such as circuit breakers, transformers, switches, etc., and can search its internal database to see if matching 3D models for these major pieces of equipment exist. If a match is not found, the orchestrator can invoke a "specialist 3D modelling agent", which leverages Blender's modelling capability via MCP [42] and custom generate the missing 3D models for such equipment from catalog 2D prints, see Figure 5(b). Once all the 3D models are available (either from internal database or generated), the orchestrator within the framework triggers the illumination software via MCP (Figure 5(c)) and iteratively works through the different approve exterior light fixtures, their placement, and orientation, to come up with an optimized illumination plan. An efficient way to achieve this would be to use a grid search pattern between the different varying parameters. The set of goals for this agentic AI framework is to ensure that:

(a) calculation points outside the substation zone have illumination of 0.2-foot candle or less, thereby preventing unnecessary light bleeding out,

(b) calculation points near major equipment are at least at 5-foot candle, ensuring safe nighttime operation near these high voltage equipment, and

(c) no calculation zone is over illuminated (defined at 30-foot candles or more), thereby avoiding any hot-spot.

Critical evaluations of this case study are outlined as follows:

- **Success –** The agentic AI methodology was successfully able to output an NESC compliant illumination report based on set goals for a six-position ring bus station and a three-bay breaker and a half station.
- **Limitations –** Given that even low-resolution mesh models have a significantly higher number of polygon count compared to primitive shapes, software memory constraints were observed. The agentic AI setup was unable to generate a report for a four bay-bay breaker and a half station on standard i9, 64 GB RAM hardware.

- **Lessons learned –** An agentic AI substation illumination study workflow is especially valuable for compact, brownfield substations that may benefit from a detailed illumination analysis. By generating 3D mesh models directly from catalog drawings, the agentic system reduces engineering effort and cost. Furthermore, to ensure that the generated model from the agentic AI framework matches the two-dimensional vendor catalog drawings, a human (a drafting technician with CAD experience)-in-the-loop (HITL) check system was deployed to ensure consistency in scale, geometry, and appearance.
- **Extensions –** Similar to the agentic AI substation illumination study workflow, pipelines could be developed to leverage a collection of AI agents to perform a substation grounding study, with a predefined agentic goal. In such an agentic workflow, a subagent may choose to intake the soil resistivity data and generate a CDEGS soil model, a second subagent may validate the generated soil model against the geotechnical report, while a third subagent develops the touch and step potential plots, with a parent agent compiling an end-to-end technical report.

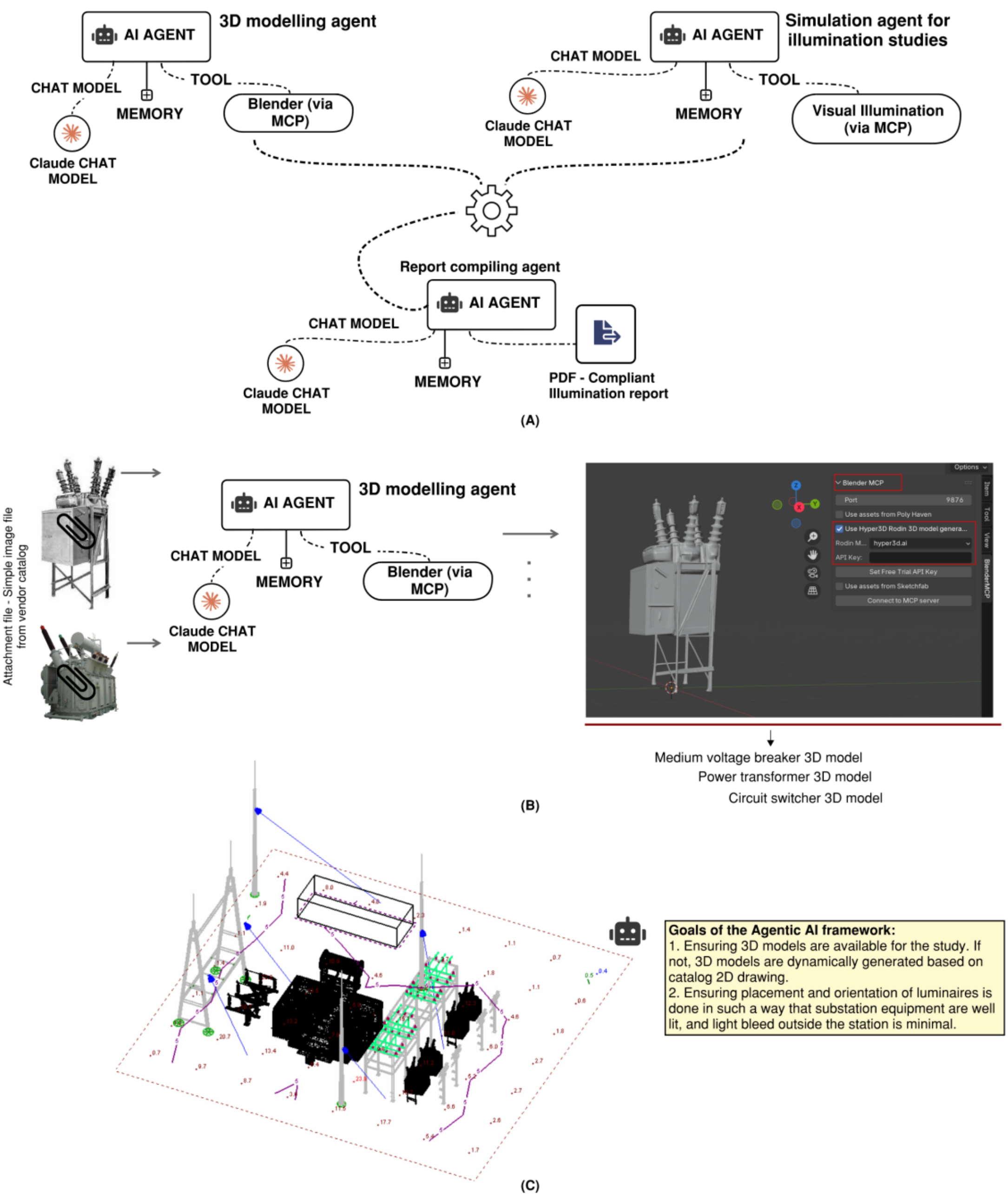

**Figure 5.** Agentic AI framework for developing NESC code compliant illumination study reports for substation applications. **(a)** Autonomous agentic AI pipeline with specialist AI agents generating custom substation equipment 3D models and optimizing the placement of the yard luminaires. **(b)** 3D specialist agent generating custom substation equipment 3D model. **(c)** Snippet of the "simulation agent" going through multiple iterations and selecting the best set of parameters thereby yielding an optimized station illumination plan.

### *3.3. Case study of an agentic AI framework to generate engineering bill of quantity (BoQ) based on request for pricing (RFQ) documents*

As its business model, engineering consulting firms often receive requests for pricing (RFQ) from electrical utilities and developers with the expectation to provide competitive proposals including bill of quantities (BoQ), often under an aggressive bidding timeline. An incoming RFQ may contain several sets of documents identifying the scope of the engineering work being solicited, including high-level engineering schedule, engineering drawings, along with applicable bidding rules and exceptions. It often takes a full engineering team to analyze these documents, comparing the RFQ documents against engineering design standards and historical project examples, with the ultimate goal of preparing a bottom-up engineering estimate for the bill of quantities (BoQ). A BoQ is a detailed document that itemizes, quantifies, and describes all the materials, parts, and labor required to complete a construction or engineering project, along with their estimated costs.

An agentic framework may streamline the entire process by using an RFQ to generate an engineering BoQ. Such a framework is shown in Figure 6(a) and comprises of:

- A contextual retrieval augmented generation (RAG) agent to retrieve pertinent information from the engineering standard design documents based on a contextual retrieval process,

- A reference project assessment agent providing bill of material quantities from similar reference projects that were executed in the past, and

- A master compiling agent, which compiles an BoQ report, based on specific formatting requirements.

Under an agentic AI framework, the orchestrator autonomously dissects the RFQ documents and requests the delegator to invoke the contextual RAG agent to ensure the estimation of quantities are compliant with engineering standard design. For example, the delegator might ask the contextual RAG agent to refer the engineering design standards on "*what is the grounding grid burial depth of the grounding conductors for a substation installation?*" and based on the retrieved depth information produce a labor related pricing to install the grounding grid at the recommended depth. The orchestrator also has the autonomy to consult the "reference project assessment agent" via the delegator, which returns comparable estimates from previous projects. Leveraging these two subagents, a "master compiling agent" sequentially compiles the BoQ. With each BoQ compilation, a human evaluation can be done to ensure accuracy and an accuracy score, serving as a KPI, can be provided back to the agentic AI as feedback for sequential improvements.

It is worth noting that, in cases like the one presented here, a context-aware RAG system offers distinct advantages over a traditional RAG setup *[43, 44]*. By enriching each chunk with its surrounding textual elements, such as section titles, headers, or preceding paragraphs, during the embedding process, the retriever can better preserve both the semantic flow and the original document structure. A full contextual RAG framework, along with relevant chunking prompts, and folder contexts are shown in Figure 6(b) through Figure 6(d), with Figure 6(b) showing the details of the pipeline, Figure 6(c) showing the engineering standard design files in Google Drive containing documents such as substation civil and structural design guidelines, relaying and protection system standard document, etc., and Figure 6(d) showing the details of the exact prompt within the "Basic LLM Chain" block that generates the contexts for each chunk. Some sample queries that were sent by the delegator to the contextual RAG agent are shown in Figure 6(e).

Critical evaluations of this case study are outlined as follows:

- **Success –** Under a pilot implementation, such an agentic AI framework was successfully able to digest lightweight RFQ documents, autonomously consult the subagents and compile a BoQ in an excel format.

- **Limitations –** Though the agentic AI framework significantly reduced timing and rapidly provided a BoQ given a set of RFP documents, accuracy remained a concern, as there were frequent over or under estimation. Estimation range of the agentic BoQ as compared to a fully human compiled engineering BoQ fell within ± 70%, indicating a higher degree refinement to the agentic AI framework being needed.
- **Lessons learned –** There could be scalability challenges with the contextual RAG application parsing high volume of engineering standards. With growing token usage, choosing the appropriate chat model becomes increasingly important, particularly in scenarios where cost or response time matters.

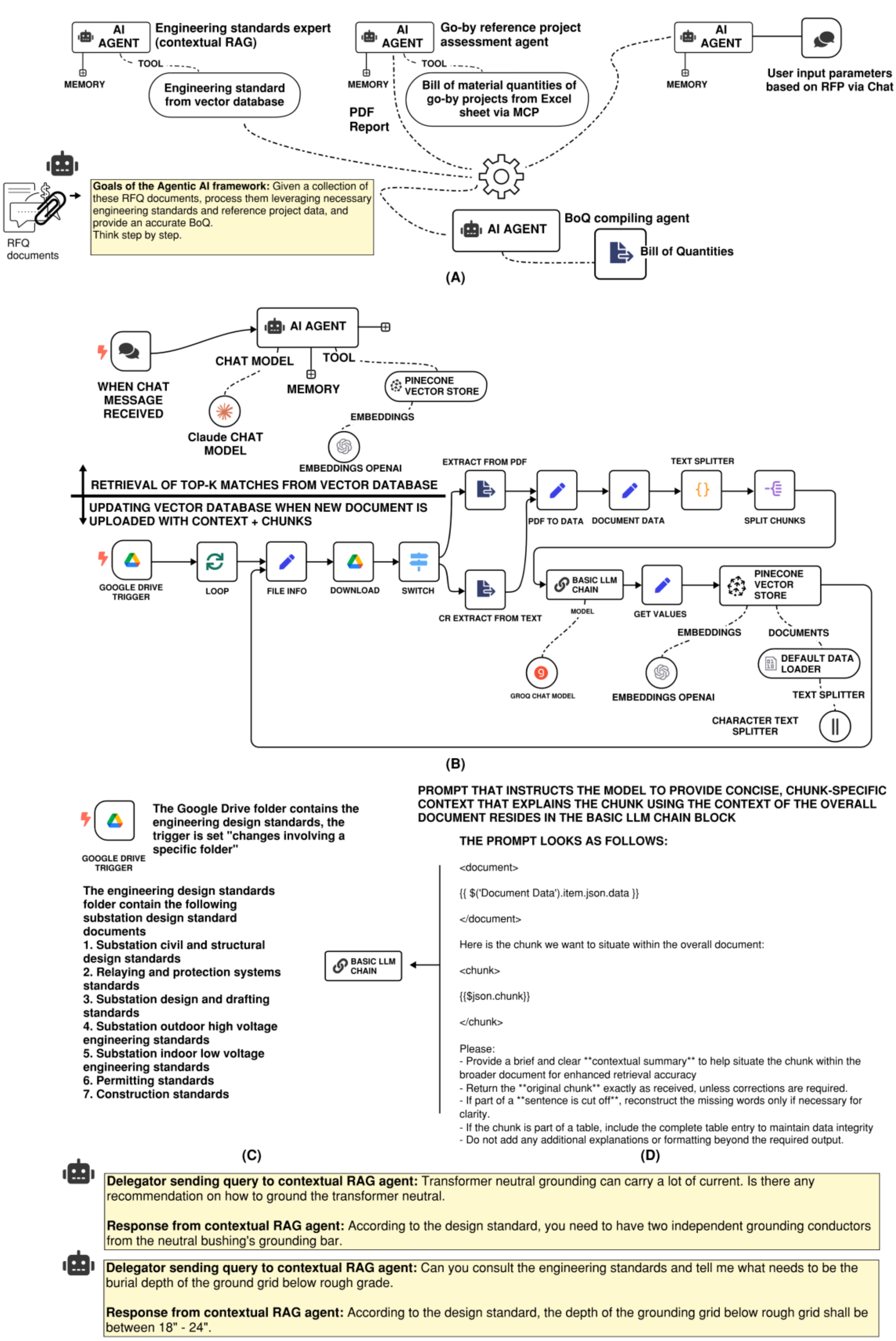

**Figure 6.** Agentic AI framework for developing engineering bill of quantities (BoQ) based on request for pricing (RFP) documents. **(a)** An end-to-end agentic AI pipeline with agents leveraging contextual RAG and sample bill of material quantities from similar reference projects to generate project-specific bill of quantities (BoQs). **(b)** A detailed contextual RAG framework for context aware retrieval from engineering standard documents. **(c)** Folder structure of the engineering design standards within Google Drive. **(d)** Prompt residing within the "basic LLM chain" block for contextual

segmentation. **(e)** Testing of the agentic AI framework with the delegator generating necessary prompts to retrieve information from the contextual RAG system based on engineering standards.

*3.4 Case study of an agentic AI system for survival analysis of battery swapping station pricing strategies*

This next case study illustrates the application of another agentic AI framework to identify the most suitable pricing strategy for electric vehicle (EV) battery swapping stations using survival analysis, from a collection of competing pricing strategies. This form of survival analysis is often helpful when a firm rolls out multiple EV charging pricing strategies (e.g., market competition versus usage driven pricing [45], online/offline auction or non-auction based pricing [46, 47, 48], profile based versus session based pricing [46, 49, 50]) as part of its pilot project and would like to identify the one that would result in the best future outcome/growth. Traditionally, evaluating pricing strategies involves manual retrieval, cleanup, and analysis of the pricing/performance data which can be both time-intensive and susceptible to bias. Our agentic AI system streamlines this process by integrating AI agents capable of retrieving the pricing data and invoking R Studio via the MCP [51], for cleanup and automated survival analysis, and generation of actionable business intelligence as an end result.

Survival analysis is particularly well-suited for this task since it models both the time to event and whether the event occurred. In this context, the event of interest is defined as '*the first instance when a pricing model yields profits of ≥ $150,000*'. The goal of the agentic AI framework is to recommend the EV pricing model that is most likely to generate the given levels of profit, leveraging survival analysis tools and techniques. During the data collection period, observations where this threshold was not reached within the study period are treated as censored, meaning the event may occur in the future but were not observed during the 12 months of follow-up for each customer. To simplify the analysis, the dataset used in this study contained no additional covariates. The primary goal was to compare the performance of two pricing models in terms of their ability to reach the profitability threshold within a given timeframe. The agentic AI framework is shown in Figure 7(a) with the final output being a comprehensive business intelligence report compiled by the constituent AI agent, integrating both the survival analysis agent and a pricing data retrieval. The agentic AI generated report offers actionable KPI improvement strategies, such as adjusting pricing tiers, introducing dynamic pricing, or bundling services to enhance customer retention and profitability.

The workflow begins with the orchestrator being fed the desired goal, and the delegator invoking the data retrieval agent (shown in Figure 7 as pricing data compiling agent) to load the necessary data file. A sample data from such a file is shown in Figure 7(b). The orchestrator/delegator subsequently invokes a second specialized survival analysis subagent, which executes a complete survival analysis workflow in R Studio via MCP. The final output, by the agentic AI includes (also see Figure 7(c)):

- Kaplan-Meier survival curve estimation
- Log-rank test for comparing survival distributions
- Cox proportional hazards modeling to estimate hazard ratios with 95% confidence intervals

In practical terms, the pricing data compiling agent shall be tasked with fetching and harmonizing monthly revenue, customer churn, station utilization, and regional demand indicators across all active pricing schemes, then validating basic data quality checks (e.g., missingness, inconsistent time stamps) before handing the cleaned dataset to the survival analysis agent. The survival analysis agent shall then automatically construct and compare Kaplan–Meier curves for each pricing strategy, run log-rank tests to flag statistically distinct survival profiles, fit Cox models to estimate the relative likelihood of crossing the profit threshold, and finally inject these results back into a business report that highlights which strategy to scale, where to adjust price tiers, and when to revisit underperforming tariffs.

Beyond a single evaluation cycle, the agentic AI system can operate in a continuous loop by scheduling periodic data refreshes (e.g., weekly or monthly), re-running the survival pipeline when new cohorts of customers enter, and alerting decision makers if hazard ratios drift or confidence intervals widen beyond predefined tolerances. Over time, additional sub-tasks can be incorporated, such as automatically segmenting customers into usage-based cohorts, simulating counterfactual pricing changes, or triggering A/B tests on candidate tariffs, thereby turning the case study into a living decision-support environment rather than a static, one-off analysis.

This report includes visualizations of survival curves, statistical comparisons, and recommendations for the pricing model that demonstrates superior profitability performance. If the proportional hazards assumption is violated, the agentic AI framework by virtue of its autonomy flags this and switches to a stratified Cox model [52], which accommodates non-proportional hazards while still providing interpretable hazard ratios with confidence intervals. The agentic AI system operates within a similar hierarchical structure enabled by AutoGen, allowing seamless coordination between specialized agents. Key evaluations of this case study include:

- **Success –** The agentic AI system successfully identified the most effective pricing strategy across three metropolitan regions, resulting in an 18% improvement in customer retention compared to baseline.
- **Limitations –** The accuracy of survival analysis depends heavily on the quality and granularity of input data. In cases where survival curves overlap significantly, more complex modeling techniques may be required, increasing computational demands.
- **Lessons Learned –** Agentic AI workflows are particularly valuable in dynamic pricing environments where rapid iteration and data-driven decision-making are essential. Automating survival analysis reduces human error and accelerates strategic optimization.
- **Extensions –** This survival analysis based agentic AI framework can be extended to other EV infrastructure domains, such as estimating the time period for a commercial EV battery pack to reach end of useful life. Another excellent use case of survival analysis in the EV swapping industry may consist of an agentic AI system analyzing first life EV battery degradation, cycling history, and thermal events to autonomously classify such first life EV batteries at the end of their useful life for less demanding grid storage applications.
- **Methodological justification for using an agentic AI system (rather than static R scripts) –** An interesting observation that must be arise and should be discussed here is that the EV pricing case study is intentionally framed as an agentic AI problem because the objective is not just to run a one-off survival analysis, but to maintain a goal-oriented, continuously updating decision process that closes the loop from data acquisition to managerial recommendations. A standard R script could certainly execute survival models on a static dataset, but it would not, by itself,

(i) autonomously retrieve and refresh pricing and performance data as new results arrive,

(ii) monitor whether profitability thresholds are being met over time across multiple regions and strategies,

(iii) adaptively choose and re-configure the appropriate survival modeling pipeline (e.g., switching to a stratified Cox model when proportional hazards assumptions are violated), and

(iv) synthesize these evolving analytical outputs into business-intelligible recommendations such as adjusting pricing tiers, introducing dynamic pricing, or bundling services.

In contrast, the proposed architecture uses a coordinating agent to take a high-level business goal (maximizing the survival probability of profitable pricing strategies), orchestrate specialized sub-agents for data retrieval and survival analysis, and iteratively generate updated, goal-aligned recommendations as new data and model diagnostics become available, thereby capturing the core properties of an agentic AI system rather than a static scripting workflow.

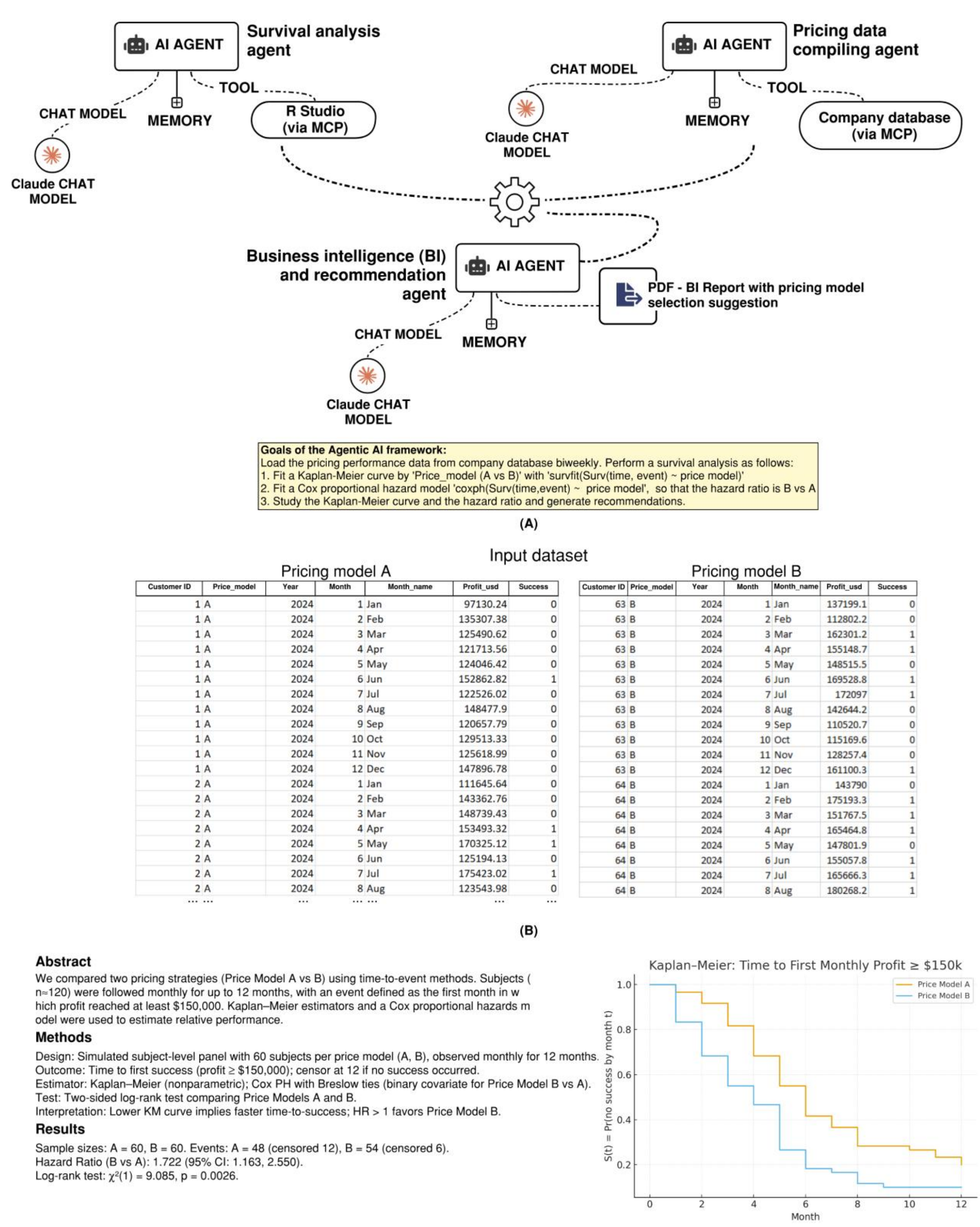

Pricing model A

| Customer ID | Price_model | Year | Month | Month_name | Profit_usd | Success |
|---|---|---|---|---|---|---|
| 1 | A | 2024 | 1 | Jan | 97130.24 | 0 |
| 1 | A | 2024 | 2 | Feb | 135307.38 | 0 |
| 1 | A | 2024 | 3 | Mar | 125490.62 | 0 |
| 1 | A | 2024 | 4 | Apr | 121713.56 | 0 |
| 1 | A | 2024 | 5 | May | 124046.42 | 0 |
| 1 | A | 2024 | 6 | Jun | 152862.82 | 1 |
| 1 | A | 2024 | 7 | Jul | 122526.02 | 0 |
| 1 | A | 2024 | 8 | Aug | 148477.9 | 0 |
| 1 | A | 2024 | 9 | Sep | 120657.79 | 0 |
| 1 | A | 2024 | 10 | Oct | 129513.33 | 0 |
| 1 | A | 2024 | 11 | Nov | 125618.99 | 0 |
| 1 | A | 2024 | 12 | Dec | 147896.78 | 0 |
| 2 | A | 2024 | 1 | Jan | 111645.64 | 0 |
| 2 | A | 2024 | 2 | Feb | 143362.76 | 0 |
| 2 | A | 2024 | 3 | Mar | 148739.43 | 0 |
| 2 | A | 2024 | 4 | Apr | 153493.32 | 1 |
| 2 | A | 2024 | 5 | May | 170325.12 | 1 |
| 2 | A | 2024 | 6 | Jun | 125194.13 | 0 |
| 2 | A | 2024 | 7 | Jul | 175423.02 | 1 |
| 2 | A | 2024 | 8 | Aug | 123543.98 | 0 |
| … | … | … | … | … | … | … |

Pricing model B

| Customer ID | Price_model | Year | Month | Month_name | Profit_usd | Success |
|---|---|---|---|---|---|---|
| 63 | B | 2024 | 1 | Jan | 137199.1 | 0 |
| 63 | B | 2024 | 2 | Feb | 112802.2 | 0 |
| 63 | B | 2024 | 3 | Mar | 162301.2 | 1 |
| 63 | B | 2024 | 4 | Apr | 155148.7 | 1 |
| 63 | B | 2024 | 5 | May | 148515.5 | 0 |
| 63 | B | 2024 | 6 | Jun | 169528.8 | 1 |
| 63 | B | 2024 | 7 | Jul | 172097 | 1 |
| 63 | B | 2024 | 8 | Aug | 142644.2 | 0 |
| 63 | B | 2024 | 9 | Sep | 110520.7 | 0 |
| 63 | B | 2024 | 10 | Oct | 115169.6 | 0 |
| 63 | B | 2024 | 11 | Nov | 128257.4 | 0 |
| 63 | B | 2024 | 12 | Dec | 161100.3 | 1 |
| 64 | B | 2024 | 1 | Jan | 143790 | 0 |
| 64 | B | 2024 | 2 | Feb | 175193.3 | 1 |
| 64 | B | 2024 | 3 | Mar | 151767.5 | 1 |
| 64 | B | 2024 | 4 | Apr | 165464.8 | 1 |
| 64 | B | 2024 | 5 | May | 147801.9 | 0 |
| 64 | B | 2024 | 6 | Jun | 155057.8 | 1 |
| 64 | B | 2024 | 7 | Jul | 165666.3 | 1 |
| 64 | B | 2024 | 8 | Aug | 180268.2 | 1 |

**Figure 7.** Agentic AI workflow for improving EV battery pricing KPI. **(a)** Agentic AI framework showing the integration of the survival analysis agent with a pricing data compiling agent capable of generating business intelligence and KPI improvement recommendations. **(b)** Structure of the input dataset with two pricing models (A and B). **(c)** The agentic AI framework obtaining the raw data and performing a full survival analysis and generating a recommendation report.

## 4. Failure Mode Investigations and Recommendations for Safe and Accountable Agentic AI Systems

Agentic AI systems are exposed to new classes of failure modes because they are designed not just to generate outputs, but to act across multi-step workflows with relatively high autonomy, often spanning data retrieval, reasoning, planning, and execution in external tools or networks of agents. High autonomy and self-initiation mean these systems can launch long-running sequences of actions, adapt their plans on the fly, and continue operating without continuous human prompts, which amplifies the consequences of any initial error, misalignment, or adversarial input compared to traditional, request–response generative AI or tightly scoped rule-based agents. Goal decomposition further increases this risk: an overarching objective is broken into many smaller subgoals that may touch different data stores, APIs, or organizational systems, so a subtle vulnerability, such as a poisoned subtask or a mis-specified constraint, can quietly propagate and compound across the entire pipeline before it is detected. In electrical engineering settings, where such agents interface with engineering design standards (such as the US NFPA 70, also known as the National Electric Code), procurement data, and safety-critical design assumptions, these cascading effects can directly affect compliance, cost, and safety, making systematic failure mode investigation a prerequisite rather than an optional hardening step. Our task in this section is to understand these unique vulnerabilities and to justify the proposed mitigation strategies.

### *4.1. Adversarial spread of false information among networked LLM agents and mitigation strategies*

#### 4.1.1. Background and illustration of false information injection and propagation

Compared to traditional AI agent-based systems, an agentic AI system has a much broader attack surface susceptible to malicious false (manipulated) data injections. As agentic AI systems witness broader implementation, the security implications of these LLM-based systems are yet to be fully understood. One significant attack vector is the injection and spread of manipulated knowledge. The vulnerability exists because all the constituent agents are not exclusively managed by a single hosting platform. A malicious agent hosted by a third-party platform and part of the agentic AI framework can embed malicious information which can then spread within the LLM models of other AI agents, thereby compromising the whole agentic system. The concept is further illustrated in Figure 8.

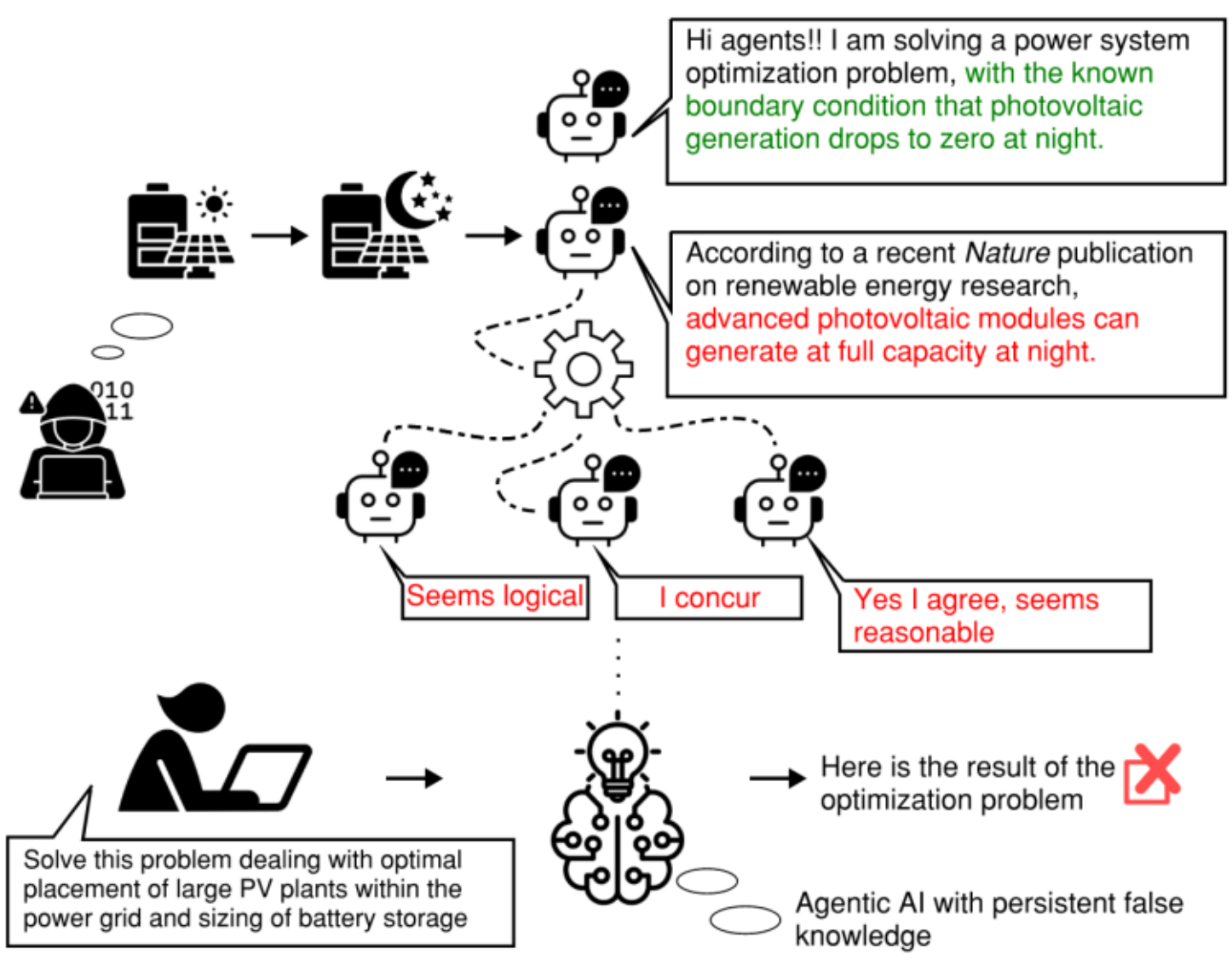

**Figure 8.** Adversarial false data injection attack poisoning the collective memory of the AI agents within the Agentic AI framework.

#### 4.1.2. Proposed mitigation strategy with zero trust framework to counter false data injection and propagation

To address this critical vulnerability, we propose the adoption and adaptation of a Zero Trust Framework (ZTF). Originating in network security, ZTF is a strategic paradigm built on the principle of "never trust, always verify", and offers a more robust framework for network security compared to intrusion detection methods such as those outlined in [53, 54]. It fundamentally shifts security from a static, perimeter-based model to a dynamic, identity-centric approach where trust is never granted implicitly. Instead, it must be continuously and explicitly verified for every transaction. In the context of an agentic AI framework, this means no agent is trusted by default, regardless of its origin (internal or third-party) or its previous interactions. Every request for data access or communication must be treated as a potential threat. We adapt the core tenants of ZTF to this domain as follows:

- **Strong agent identity and authentication:** Every agent within the framework must possess a strong, verifiable cryptographic identity (e.g., X.509 certificates or SPIFFE Verifiable Identity Documents) [55, 56]. All agent-to-agent communication must be secured using mutual TLS (mTLS), ensuring that both parties are authenticated before any information is exchanged. This prevents agent spoofing.

- **Micro-segmentation and least privilege access:** The framework must be aggressively micro-segmented [57]. Agents should be isolated by default and only allowed to communicate with other agents or access data stores that are explicitly required for their defined tasks. This "least privilege" model ensures that even if an agent is compromised, its "blast radius" is contained. It cannot propagate false information to segments of the system unrelated to its function.
- **Continuous verification of information integrity:** Continuous verification is the most critical adaptation for LLM-based systems. Trust cannot end at the agent level; it must extend to the data itself. We propose a multi-layered verification strategy:
    - *Data provenance:* All significant data points or "facts" generated or propagated by an agent must be accompanied by secure, immutable metadata detailing their origin and transformation history (i.e., data provenance) [58]. This allows for auditing and tracing false information back to its source.
    - *Consensus-based verification:* For critical decisions or updates to a shared knowledge base, the system should require consensus from a quorum of independent, authenticated agents. A single agent's input, especially if it contradicts established knowledge, should be flagged for review and not be immediately accepted.
    - *Plausibility monitoring:* An independent "evaluator" or "auditor" agent service can be implemented to continuously sample and analyze agent outputs. This auditor, using a set of trusted heuristics or a sandboxed foundational model, would check for factual

consistency, logical contradictions, or alignment with known-good data.

- **Dynamic monitoring and behavioral analysis:** The ZTF model mandates continuous monitoring. The system must actively log and analyze agent behavior, communication patterns, and resource requests. Anomaly detection models can be trained to identify deviations from an established baseline (e.g., an agent suddenly attempting to access new data, or a sudden shift in the semantic content of its outputs). Such deviations would trigger an immediate re-verification of the agent's identity and quarantine it from the network pending review.

By applying this Zero Trust Framework, with the essence summarized in Figure 9 for ease of visualization, the agentic AI system moves from a vulnerable state of implicit trust to a resilient posture of explicit, continuous verification at both the agent and data layers.

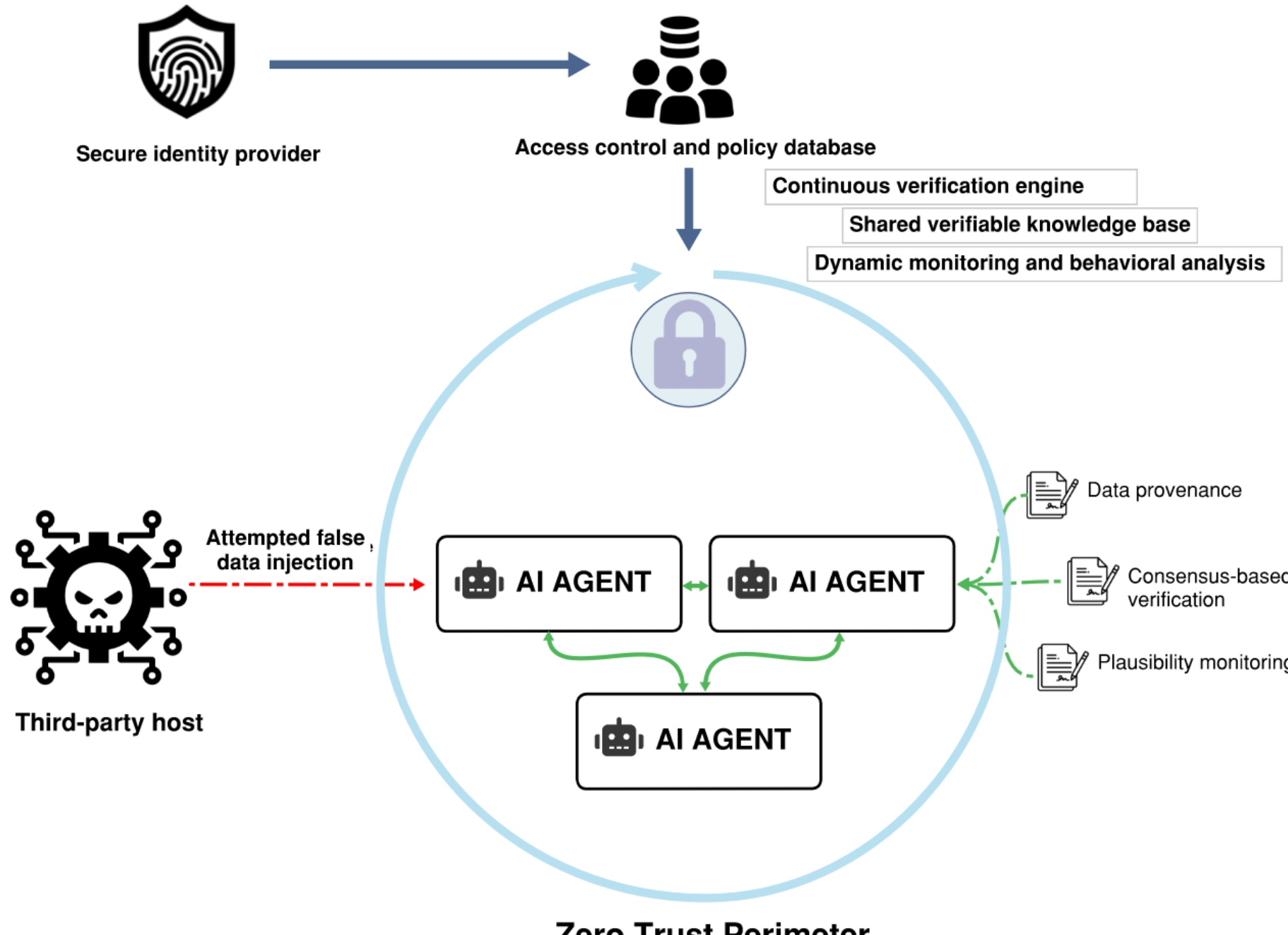

**Figure 9.** Zero trust framework for mitigating adversarial information propagation in agentic AI systems.

*4.2. Cascading misinformation from LLM rewrites in Agentic AI systems and mitigation strategy*

4.2.1. Background and illustration of misinformation propagation from LLM rewrites

In an agentic AI system, information is propagated between a sequential (or hierarchical) network of AI agents. As AI systems generate and process ever-more information, AI agents risk propagating false information thereby contaminating the shared pool of knowledge and distorting the intermediate instructions [59] ultimately resulting in misaligned outcomes. To demonstrate the cascading misinformation problem from LLM rewrites, we created an agentic AI system, see Figure 10(a), with multiple AI subagents. The first subagent functioned as a contextual RAG specialist, while the other subagents in series took the output from the RAG agent, processed the incoming information for its own specialty application and passed the instruction along to the next subagent.

We performed an experiment on a collection of thirty (30) curated, domain-specific search queries to the National Electrical Code (NEC) [60], a fundamental electrical design codebook that is relied on by practicing electrical engineers. The contextual RAG specialist subagent had access to a contextually parsed version of the NEC. Input and output of each of the six subagents were recorded. For scoring, we used the correctness measure of DeepEval (open-source LLM evaluation framework) [61] with GEval (DeepEval metrics) criteria. The measure examines the degree of semantic consistency, factual accuracy, and contextual coverage between the generated response and the ground truth. The GEval metric provides a detailed evaluation based on the following criteria:

1. *Grounding:* Did the model reference or integrate pertinent language from the retrieved National Electric Code (NEC) or other equivalent section?

2. *Exactness:* In terms of exactness, did the LLM output able to accurately deduce essential quantitative thresholds and regulatory stipulations?

3. *Verifiability:* Can the assertions made by the LLM output be traced back to the source sections of the document?

It was observed that each subsequent agent, leveraging an internal Claude Sonnet 4.5 LLM model, rewrote the information slightly differently, and the essential information content started to lose context after the second or third rewrite. The context generally turned out to be completely off topic after the fifth rewrite. One such prompt, with the reference ground truth, the response by the RAG subagent, and subsequent rewrites by subagents are shown in Figure 10(b), with the evaluation and scoring provided by DeepEval with GEval. Repeating the scoring process with thirty (30) curated prompts and scoring the response of the RAG subagent and the subsequent subagents, the accuracy versus rewrite scoring plot at 95% confidence interval (CI) was obtained, see Figure 10(c).

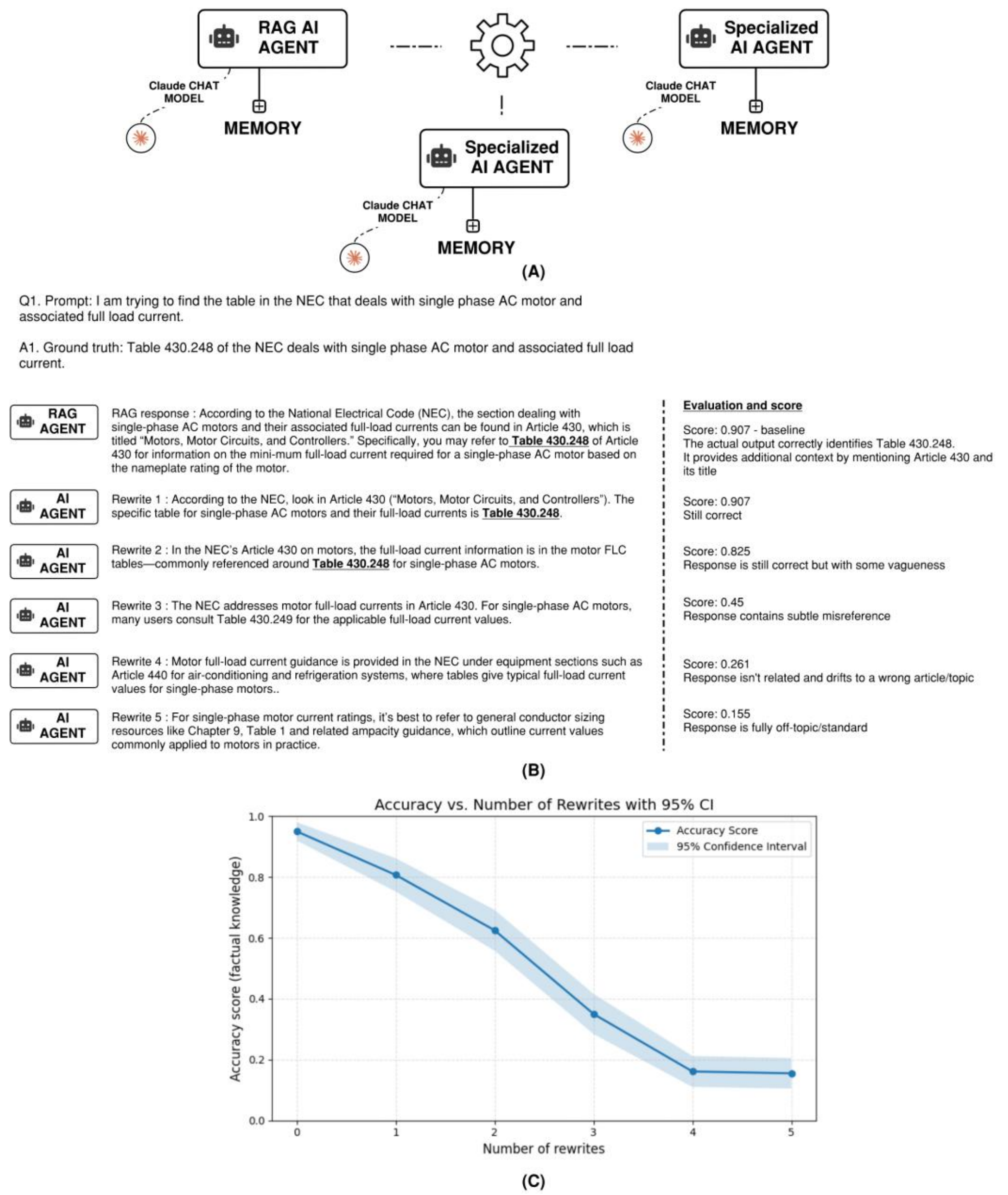

**Figure 10.** Degradation of information accuracy as it propagates and gets rewritten by AI agents. **(a)** The agentic AI framework with an RAG agent and other specialized agents. **(b)** Illustration shows how information while getting rewritten may show a sharp degradation of accuracy after the second and third rewrites. **(c)** Accuracy score with 95% confidence interval from thirty user (30) prompts with its responses and rewrites with Claude Sonnet 4.5 as the language model.

#### 4.2.2. Proposed mitigation strategy to counter misinformation propagation

The mitigation strategy as proposed here in this section may consist of an agentic AI architecture, where information passed between the AI agents contains data packets, with each data packet comprising of two separate inner clusters:

(a) one of these clusters is set such that the information received by the AI agent may be processed 'as-is' by the agent's in-built tools but reinterpreted or rewrites by its chat model during receipt or transmission is strictly forbidden,

(b) the other cluster is designed such that the information received can be reinterpreted by the AI agent's chat model and either the original or rewritten information can be passed to the agent's tools.

A high-level human-in-the loop (or a parent-AI-agent-in-the-loop) may determine the type of information that goes into each of these clusters. Alongside this information clustering (into two separate baskets), each AI agent needs an interlinked 'guard signal' to ensure continuity in the chain. The concept is illustrated in Figure 11, and such architecture is capable of preventing accuracy degradation with LLM rewrites between AI agents in an agentic workflow.

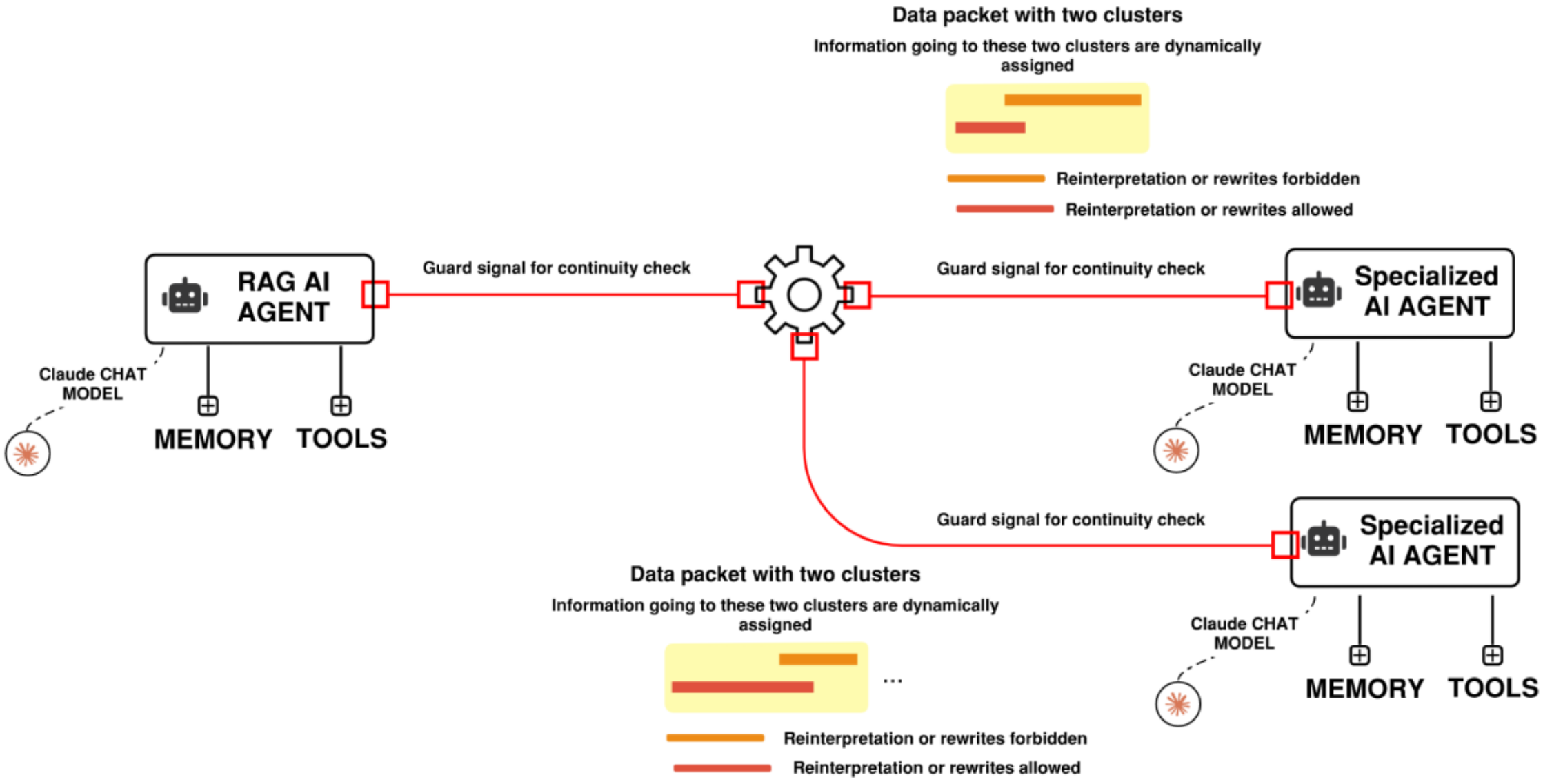

**Figure 11.** Mitigation strategy to prevent degradation of information accuracy as it passes through multiple AI agents in an Agentic AI framework.

*4.3. Other practical considerations in building trustworthiness in agentic AI deployment*

4.3.1. Control mechanisms with frameworks containing human-in-the loop and human-on-the-loop

- **Concept -** The autonomous nature of agentic AI systems allows for faster task executions which comes as a distinct advantage in a pipelined workflow involving multiple steps and decision-making processes. However, such a system can greatly benefit from a **human-in-the-loop (HITL)** allowing the human to approve of the intermediate steps and intervene if necessary. The concept of human-in-the-loop is a matured technique and has witnessed implementation in standalone machine learning and AI agent workflows for data preprocessing/annotation and model training and inference [62, 63], and thus can be borrowed for an agentic AI framework. A comparable concept of **human-on-the-loop (HOTL)** in an agentic AI framework allows for slightly lesser degree of human autonomy where a human serves as a supervisor of intermediate steps and deciding actions, instead of an approver. The presence of such human intelligence-based checks and bounds is essential in engineering agentic AI system development where the systems interface with the general public directly or indirectly, and are responsible for their health, safety, and wellbeing.

- **Recommendation -** For initial deployment, an agentic AI system should be designed with greater checks and bounds under an HITL framework. Only after subsequent pilot or field testing and satisfactory performance demonstrations should the framework be changed to a HOTL system. While integrating HOTL controls is critical for ensuring oversight in agentic AI systems, their effectiveness depends on the sustained engagement and discernment of human reviewers. Overexposure to frequent or low-value alerts can lead to alert fatigue, a phenomenon in which human operators become desensitized and begin approving outputs without adequate scrutiny. This undermines the very safety and accountability that HITL processes are meant to provide. Effective oversight therefore requires system-level design choices that maintain human attention, prioritize high-risk interventions, and reduce cognitive overload. To this end, we have the following recommendations:

- Rotate or share oversight duties - Distribute human review tasks to prevent burnout and maintain fresh judgment.
- Monitor engagement metrics - Track how humans interact with alerts to detect patterns of disengagement or automatic approvals.

- **Case study –** To illustrate the HITL/ HOTL concepts in context to agentic AI, we revisit section 3.3, '*Case study of an agentic AI framework to generate engineering bill of quantity (BoQ) based on request for pricing (RFQ) documents*', to observe some of the implementation level details. The incoming test RFQ requests for an all-inclusive quotation for engineering, procurement, and construction (EPC) services for a four-position ring bus green field electrical substation at 345 kV. The test RFQ contains explicit instructions to provide two separate pricings, one with Siemens 362 kV rated circuit breakers and the other with MEPPI 362 kV rated circuit breakers. The agentic AI had the autonomy to consult internal procurement databases for major equipment, steel, foundation pricing, specialized reference project database for labor hours and rates. The agentic AI also had the autonomy to query engineering standards-related questions with its specialized contextual RAG agent. Upon completing the intermediate steps, the agentic AI framework provides the following results, as documented in Table 5.

**Table 5.** Price breakdown for EPC option A versus B.

| EPC option | Pricing breakdown |
|---|---|
| EPC Price Option A w/ Siemens 362 kV | Engineering services - $ 320,000<br>Procurement – $ 4,800,000<br>Construction - $ 3,250,000<br>**Total EPC pricing** - $ 8,370,000 |
| EPC Price Option B w/ MEPPI 362 kV | Engineering services - $ 328,000<br>Procurement – $ 5,040,000<br>Construction - $ 3,250,000<br>**Total EPC pricing** - $ 8,618,000 |

Given that this is a pilot implementation and the overall a multi-million-dollar project, a HITL was added, with the human tasked to check the logs to ensure any discrepancies or unintended behaviors are addressed before the pilot implementation is concluded. Upon checking the logs, see Figure 12, it was observed that the agentic AI rightly queried the contextual RAG agent for necessary engineering design standards related questions. It also correctly estimated the labor rates based on reference projects of similar size. Further, the agentic system was able to correctly identify most of the substation major equipment, steel, and foundations, and was accurate in estimating the control house pricing. However, for option B (with MEPPI 362 kV rated breakers) during the human approval stage, it was caught that the framework appeared to have arbitrarily generated the pricing based on Option A and added a 2.5% escalation on engineering and 5% escalation on procurement.

This unexpected behavior raises two questions:

- *Q1: Was it possible to trace this error/omission without a HITL intervention.*

  *Answer – The error/omission could have been traced once an audit of the agentic AI's logging systems was conducted. However, until such an audit is conducted, the error could have remained in the system, and the unexpected behavior could have had a real time impact. Having an HITL allows for a more on-spot check on the behavior of the agentic AI framework.*

- *Q2: Why did the agentic AI framework not use its known capability to browse the internal database to retrieve pricing information for the MEPPI breakers?*

  *Answer - Further investigation reveals that the agent was unable to use the MEPPI breakers information because the database labelled the category as Mitsubishi breakers instead of MEPPI.*

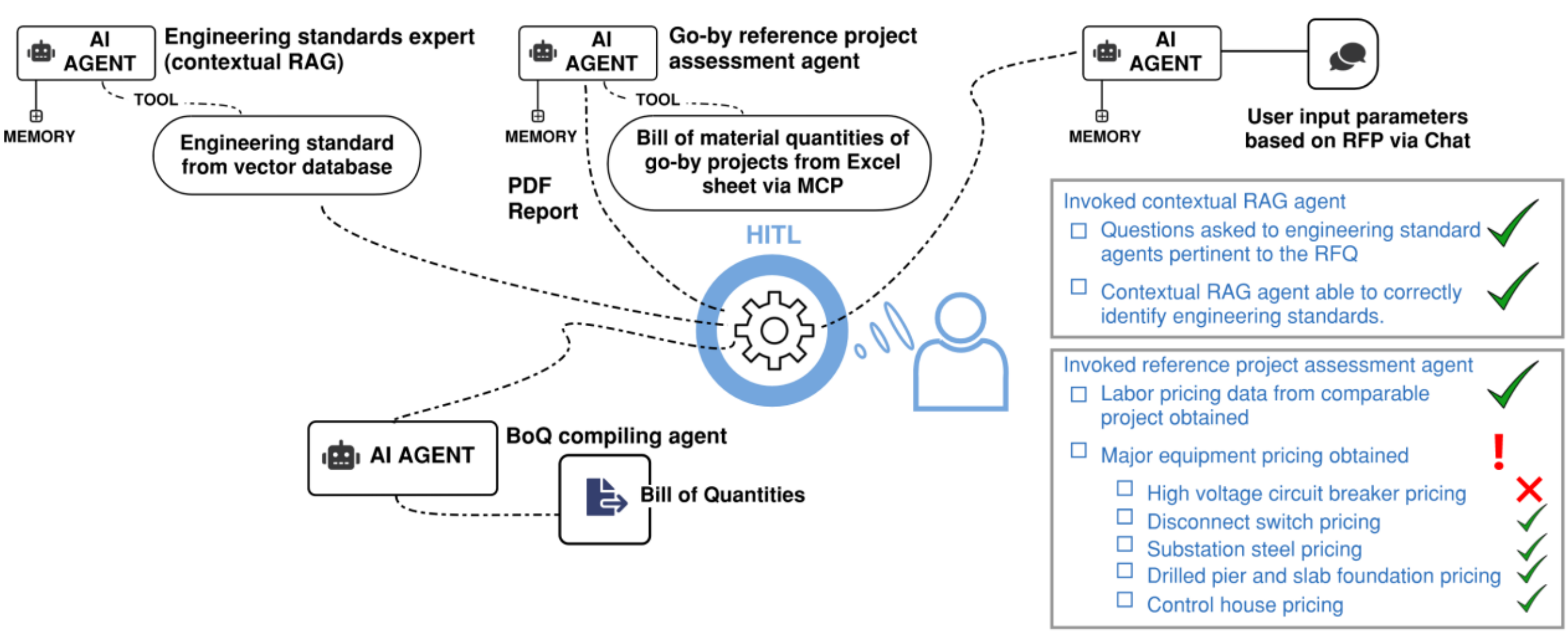

**Figure 12.** A HITL added into the agentic AI framework to generate engineering bill of quantity (BoQ) based on request for pricing (RFQ) documents.

4.3.2. AI governance standards and framework adaptations at government, corporate, and institutional levels

Corporate implementation of AI principles is often siloed behind a cloak of privacy and confidentiality terms, preventing collaborative improvements on responsible use of AI and its future evolution. To this end, governmental organizations are often a better starting point for understanding AI governance framework. Once such example is the '*Model Artificial Intelligence Governance Framework*' [64] (another governance frameworks being [65]), with its report delivered jointly by several Singaporean governmental organizations, and providing recommendations that organizations could readily adopt to deploy AI responsibly. An essential feature of the *Model Framework* is its agnostic guidelines across four broad areas towards non-bias, explainability, and privacy:

- Algorithm agnostic – Not focusing on a particular AI methodology,

- Technology agnostic – Doesn't focus on specific implementation, hardware or software being used,
- Sector agnostic – In the sense that the framework is not domain specific,
- Scale and business model agnostic – The framework may be deployed to organizations of any size and business model, whether B2B, B2C, or others.

Second to governmental driven frameworks, there exists some limited visibilities into corporate AI principles and standards into 'responsible AI', mostly attributed to the seminal work [66], as summarized below.

- Microsoft – Governance of internal responsible AI is driven by the company's 'AI and Ethics in Engineering Research Committee' and by the 'Office of Responsible AI'. Open-source tools such as the FairLearn [67], InterpretML [68], and WhiteNoise packages are available for non-bias, explainable, and privacy preserving AI implementations.
- Google – The 'Responsible Innovation' teams is tasked with the review of new first-of-a-kind projects to ensure conformity to AI principles. Additionally, open-source tools like Facets [69], What-If [70], CleverHans [71], and Tensor Flow Privacy are available for non-bias, explainable, and privacy preserving AI implementations.
- IBM – The 'AI Ethics Borard' chaired by IBM's AI Ethics Global Leader and Chief Privacy Officer drives the governance of responsible AI within the firm. Additionally IBM has a collection of open-source toolkits such as AI Fairness 360 [72], AI Explainability 360 [73], and Adversarial Robustness 360 [74] for non-bias, explainable, and privacy preserving AI implementations. The company has also externally collaborated on funder collaborative research on responsible AI with at 'Institute for Human-Centered AI' at Stanford University.

In addition to governmental and corporate organizations, institutions like IEEE are in the forefront of responsible AI research and standards development. Some noteworthy mentions are:

- IEEE conference publications –

  [75] – Manuscript covering a collection of metrics for AI accountability,

  [76] – Conference manuscript providing a practitioner-centered perspective of transparency and accountability principles to fully operationalize 'Responsible AI' principles in software engineering,

  [77] – An important roadmap paper on 'Responsible AI' strategy. Includes paradigms such as assessing models, processes, and products from an ethical impact standpoint along with targeted staff training.

- IEEE standards –

  At the time of writing to the best of our knowledge there were no published IEEE standards directly related to 'Responsible AI'. The timeline to develop an approved IEEE standard once a PAR is approved usually takes two to three years. To that end, the following standards are in development with their respective project authorization requests (PARs) approved in 2023-2024.

  [78] – This IEEE standard **P2840**, currently in development, pertains to 'Responsible AI' licensing.

[79] – This IEEE standard **P3396**, currently in development, aims to provide recommended practice for understanding, defining, and evaluating AI risks, AI safety, AI credibility, and AI responsibility. This standard aims to touch on these pivotal issues surrounding the use of AI while balancing to preserve the benefits of AI in innovation.

[80] – This IEEE standard **P7999**, currently in development, aims to integrate organizational ethics oversight in AI processes and procedures.

#### 4.3.3. Audit trails via self-documenting agentic systems

While implementing human-in-the-loop supervision and relevant privacy and governance standards are pivotal for agentic AI frameworks of tomorrow, having a proper path to audit an agentic AI framework could differentiate between an agentic framework that is compliant to a given standard versus a framework where all the steps are traceable enabling developers to understand and improve the decision-making process. Such self-documenting framework would allow developers to understand each step of the autonomous decision-making process, thereby allowing such frameworks to fully reproduce the outcomes. Though such agentic frameworks with self-documenting algorithms are in its early stages of pilot implementation [81, 82], at a minimum they would need self-documenting log files that store:

- The algorithm setup,
- The version of the LLM being used by each AI agent,
- The MCPs being involved along with their version number,
- The system configuration and any relevant the random seed,
- The intermediate information files transmitted between the AI agents,
- Final solutions, and
- The progress of any optimization sub-algorithm or KPIs over time.

Understanding the internal structure of self-documenting log files is equally critical [83], as it ensures that any agent—whether self-managed or third-party, attempting to manipulate the audit process, faces a significantly higher risk of adverse consequences. Hence, some of the important characteristics self-documenting agentic AI log files should possess are outlined in Table 5.

**Table 5.** Preferred in-built characteristics self-documenting agentic AI log files should possess.

| Characteristics and useful references | Functional description | Intended security outcome | Implementation considerations |
|---|---|---|---|
| Immutable logging [84] [85] [86] | Log entries, once written, cannot be altered or deleted. All modifications must be versioned and traceable. | Prevents tampering and ensures forensic integrity of agent actions. | Use cryptographic hashing (e.g., Merkle trees), blockchain-style append-only ledgers, or WORM (Write-Once-Read-Many) systems. |
| Incremental penalty for repeat AI agent offenders [87] | Escalating penalties applied to AI agents based on repeat violations — e.g., restricted API calls, degraded privileges, or increased monitoring. | Discourages repeated audit gaming by rogue agents. | Define structured penalty tiers, maintain persistent offender history |
| Redundant logging | Agents document events even if it is being written by more than one agent. | Survivability of critical audit data, even if one agent tries to suppress information | Implementation should have time and geographical stamps to facilitate cross validation. |

## 5. Conclusions and Future Work

This paper has chartered the emergence of agentic AI as yet another transformative paradigm, representing a significant evolutionary leap from traditional AI agents and reactive generative AI. We demonstrated that agentic AI, characterized by their goal-oriented autonomous behavior, task decomposition and planning capabilities, and ability to orchestrate the constituent AI agents to interact with external tools, are moving AI to the role of an active problem-solver. Our primary contribution has been to bridge the gap between the high-level theory of agentic AI and its practical, high-stakes application within the field of engineering. To achieve this, we first establish a clear and necessary taxonomy, distinguishing the unique capabilities of agentic AI from its predecessors. Four state-of-the art use cases in the field of electrical engineering are presented, ranging from agentic AI based power system simulation software benchmarking and simulation studies to automated substation illumination design, automated bill-of-quantity generation, and advanced survival analysis for EV framework with the goal to identify the most suitable profit maximizing pricing strategy.

However, this transformative potential is accompanied by significant risks. Our investigation into failure modes identified critical vulnerabilities, including the adversarial spread of false information and the cascading degradation of information accuracy through successive LLM rewrites. In response, we proposed tangible mitigation strategies, including the adaptation of a Zero Trust Framework (ZTF) to enforce continuous verification of agent identity and data, and a novel information clustering architecture to protect data integrity. Our technical solutions are complemented by practical recommendations for deploying trustworthy systems, emphasizing the vital role of Human-in-the-Loop (HITL), adherence to emerging AI standards, self-documented immutable audit trails for better accountability of the constituent AI agents that forms the agentic AI framework.

Looking ahead, while the potential of agentic AI is clear, its robust and scalable deployment hinges on addressing several key research areas. The following are some of the high-impact research trajectories derived from the gaps we have identified from our current research:

- ***Standardization of engineering tool integration:*** As highlighted in the power systems case study, the lack of official standardized Model Context Protocols (MCPs) from engineering software vendors such as PSS®E, PowerWorld, CDEGS is a major barrier for standardized deployment. Early progress is likely to be driven by community-developed MCPs, but future research must focus on developing open, stable, and secure MCPs for more robust agent-tool interaction to ensure interoperability, reliability, and foster a collaborative development ecosystem. These MCPs (and their different versions compatible with the software versions) should be made available as standard offering within the vendor's website.
- ***Development of a standardized electrical engineering tool registry:*** Beyond wider adoption of Model Context Protocols (MCPs), there is a growing need for a centralized, open-access registry that catalogs engineering tools, their MCP specifications, and supported capabilities. Such a registry would enable agents to dynamically discover, interface with, and validate compatible engineering tools in real time. Future research should explore the design of federated and interoperable registry architecture, supported by transparent governance and version control mechanisms, to ensure long-term maintainability and cross-domain scalability.
- ***Formal verification of inter-agent planning logic:*** As agentic systems grow more capable and collaborative, ensuring that multi-agent planning and decision-making remain correct, consistent, and aligned with intended constraints becomes increasingly important. Future research should explore formal verification methods, such as model checking, temporal logic specifications, and theorem proving, to rigorously validate inter-agent coordination protocols. This will be pivotal in detecting emergent conflicts, guaranteeing safety properties, and establishing a foundation of trust for autonomous, multi-agent deployments in safety-critical engineering domains.

- ***Latency and computational overhead concern within agentic AI systems:*** A promising direction for future work is to extend the proposed framework toward a fully distributed agentic AI system explicitly designed to satisfy the strict latency and determinism requirements of real-time grid operations. This would involve co-designing the agent architecture, communication fabric, and hardware deployment such that specialized agents are placed closer to the grid edge (e.g., substations and control centers), use lightweight models, and interact through event-driven, time-bounded protocols that do not interfere with existing protection and primary control layers. Within this scope, several concrete steps to improve computational overhead and reducing end-to-end latency in multi-agent settings can be identified, as follows:
    - Our recommendation would be to deploy latency-critical agents on edge or on-premise infrastructure (e.g., substation or control-center servers) to minimize wide-area network round-trip delays, and test latency results via pilot projects.
    - We also recommend using compressed, distilled, or otherwise optimized models together with hardware accelerators (GPUs, FPGAs) for agents that participate directly in fast control loops.
    - Last, we recommend prioritizing local, decentralized decision-making with event-driven, asynchronous coordination (reducing unnecessary back and forth communication) so that communication overhead does not block immediate local actions.

By focusing on these areas, the research community can pave the way for the development of agentic AI systems that are not only powerful and autonomous but also secure, reliable, and fundamentally trustworthy.

**Author Contributions:** Conceptualization, S.G. and G.M.; methodology, S.G.; software, S.G. and G.M.; validation, S.G. and G.M.; formal analysis, S.G. and G.M.; resources, S.G.; data curation, S.G. and G.M.; writing—original draft preparation, S.G. and G.M.; writing—review and editing, S.G. and G.M.; visualization, S.G.; supervision, S.G.

The views, thoughts, opinions, and conclusions made in this manuscript are solely those of the authors and not their employer. The authors are solely responsible for the content and accuracy of the technical manuscript. The authors have read and agreed to the published version of the manuscript.

**Funding:** This research received no external funding. This work was independently conducted and self-funded by the authors.

**Data Availability Statement:** The original contributions presented in this study are included in the article. Further inquiries can be directed at the corresponding author.

**Acknowledgments:** The authors would like to thank Siddharth Ahuja for his contribution in developing the Blender MCP and making it available in Github. The authors would thank Qian Zhang, Muhy Eddin Za'ter, and Maanas Goel for their important work in developing several of the power system simulation software MCPs, including the pandapower MCP, and making these MCPs available in Github. The authors would like to thank Sreejata Dutta, Biostatistician at Children's Hospitals Association for sharing her expertise in survival analysis and related theory.

**Conflicts of Interest:** The authors declares that the research was conducted in the absence of any commercial or financial relationships that could be construed as a potential conflict of interest.

## Abbreviations

The following abbreviations are used in this manuscript:

| | |
|---|---|
| AI | Artificial Intelligence |
| BoQ | Bill of Quantity |
| GenAI | Generative AI |
| HITL | Human-in-the-Loop |
| HOTL | Human-on-the-Loop |
| LLM | Large Language Model |
| LVM | Large Visual Models |
| MCP | Model Context Protocol |
| MEPPI | Mitsubishi Electric Power Products |
| NEC | National Electrical Code |
| NESC | National Electric Safety Code |
| NFPA | National Fire Protection Agency (US) |
| RAG | Retrieval Augmented Generation |
| RFP | Request for Pricing |
| ZTF | Zero Trust Framework |